%% file: paper.tex
\documentclass[12pt]{article}

\usepackage{amsmath}
\usepackage{amsfonts}
\usepackage{amssymb}

\usepackage{graphicx}
\usepackage{epsfig}
\usepackage{epsf}

\allowdisplaybreaks

\begin{document}

\author{A. Nehme}

\title{Splitting Strong and Electromagnetic Interactions in $K_{\ell 4}$ Decays~\footnote{This work is dedicated to my son}}
\maketitle

\begin{center}
27 rue du Four de la Terre \\
F-84000 Avignon, France
\end{center}

\begin{center}
\texttt{miryama.nehme@wanadoo.fr}
\end{center}

\begin{abstract}
We recently considered $K_{\ell 4}$ decays in the framework of chiral perturbation theory based on the effective Lagrangian including mesons, photons, and leptons. There, we published analytic one-loop-level expressions for form factors $f$ and $g$ corresponding to the mixed process, $K^0\rightarrow\pi^0\pi^-\ell^+\nu_{\ell}$. We propose here a possible splitting between strong and electromagnetic parts allowing analytic (and numerical) evaluation of Isospin breaking corrections. The latter are sensitive to the infrared divergence subtraction scheme and are sizeable near the $\pi\pi$ production threshold. Our results should be used for the extraction of the $P$-wave iso-vector $\pi\pi$ phase shift from the outgoing data of the currently running KTeV experiment at FNAL.
\end{abstract}

\textbf{keywords:} Electromagnetic Corrections, Kaon Semileptonic Decay, Form Factors, Pion Pion Phase Shifts, Chiral Perturbation Theory.

\pagebreak

\tableofcontents

\input{introduction}

\input{section_1}

\input{section_2}

\input{section_3}

\input{section_6}

\input{conclusion}

\begin{flushleft}
\textit{\textbf{Acknowledgements}}
\end{flushleft}
This work was entirely done at home. I am grateful to my wife,
Miryam Nehme, for patience and devotion to Physics advancing. In
this optics, she offered to me the computer, the Linux station,
the internet connection and the MATHEMATICA program in order to
achieve the present work.

\appendix

\input{appendix}

\bibliographystyle{unsrt}

\bibliography{paper}

\begin{figure}[p]
\epsfxsize14cm \centerline{\epsffile{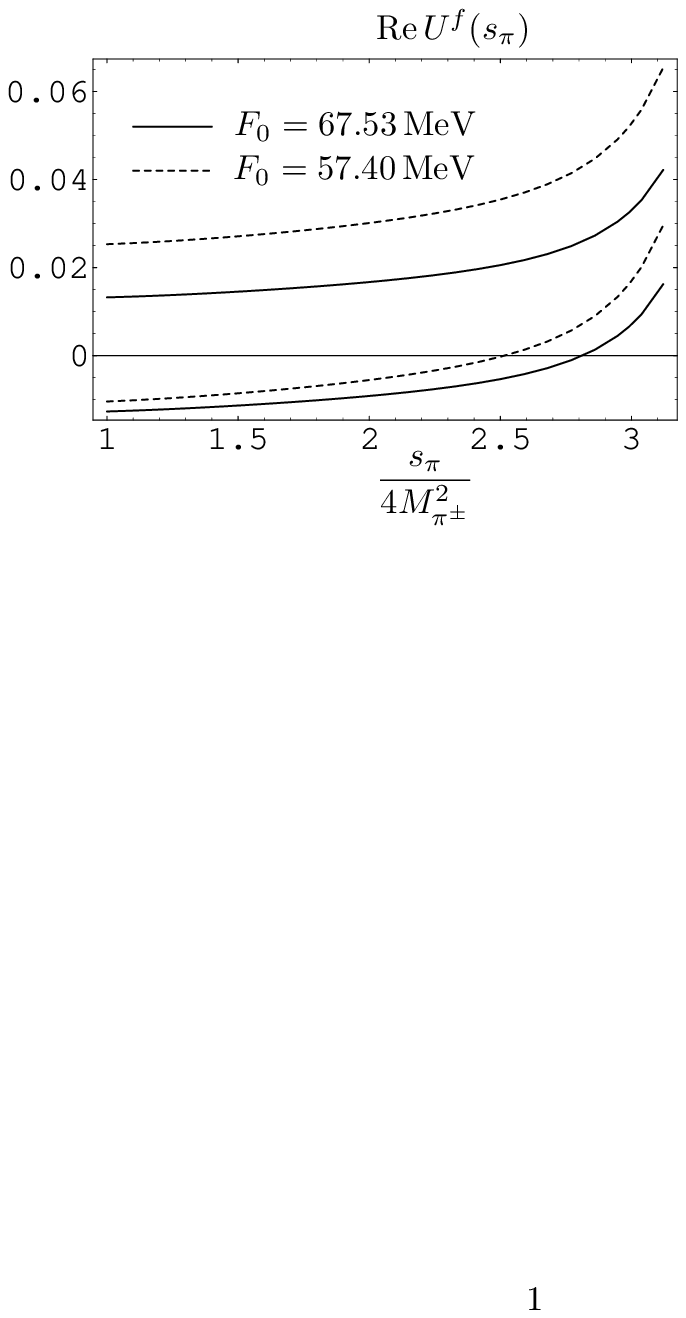}}
\caption{\label{fig:f-module_electron} The real part of the first
term in the partial wave expansion for $f$ form factor under the
assumption, $s_{\ell}=m_{\ell}^2=m_{\mathrm{e}}^2$. The error band
comes exclusively from the uncertainty in the determination of
low-energy constants and has been developed in quadrature.}
\end{figure}

\begin{figure}[p]
\epsfxsize14cm \centerline{\epsffile{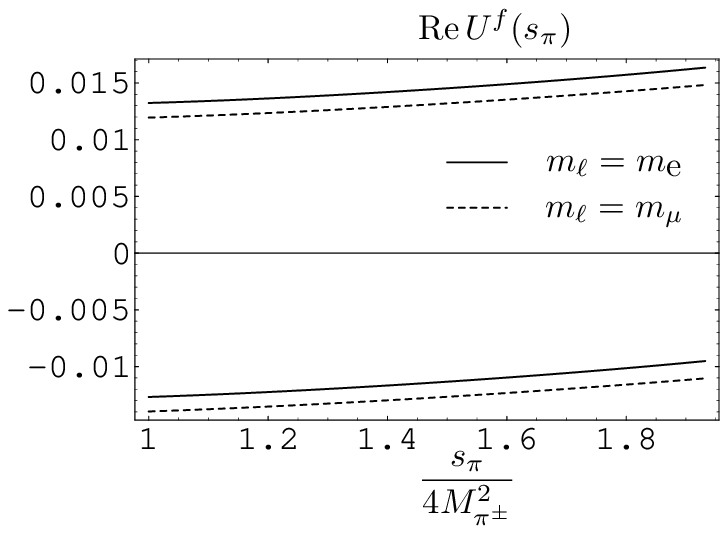}}
\caption{\label{fig:f-module_mass} The real part of the first term
in the partial wave expansion for $f$ form factor under the
assumptions, $s_{\ell}=m_{\ell}^2$, $F_0=67.53$ MeV. The error
band comes exclusively from the uncertainty in the determination
of low-energy constants and has been developed in quadrature.}
\end{figure}

\begin{figure}[p]
\epsfxsize14cm \centerline{\epsffile{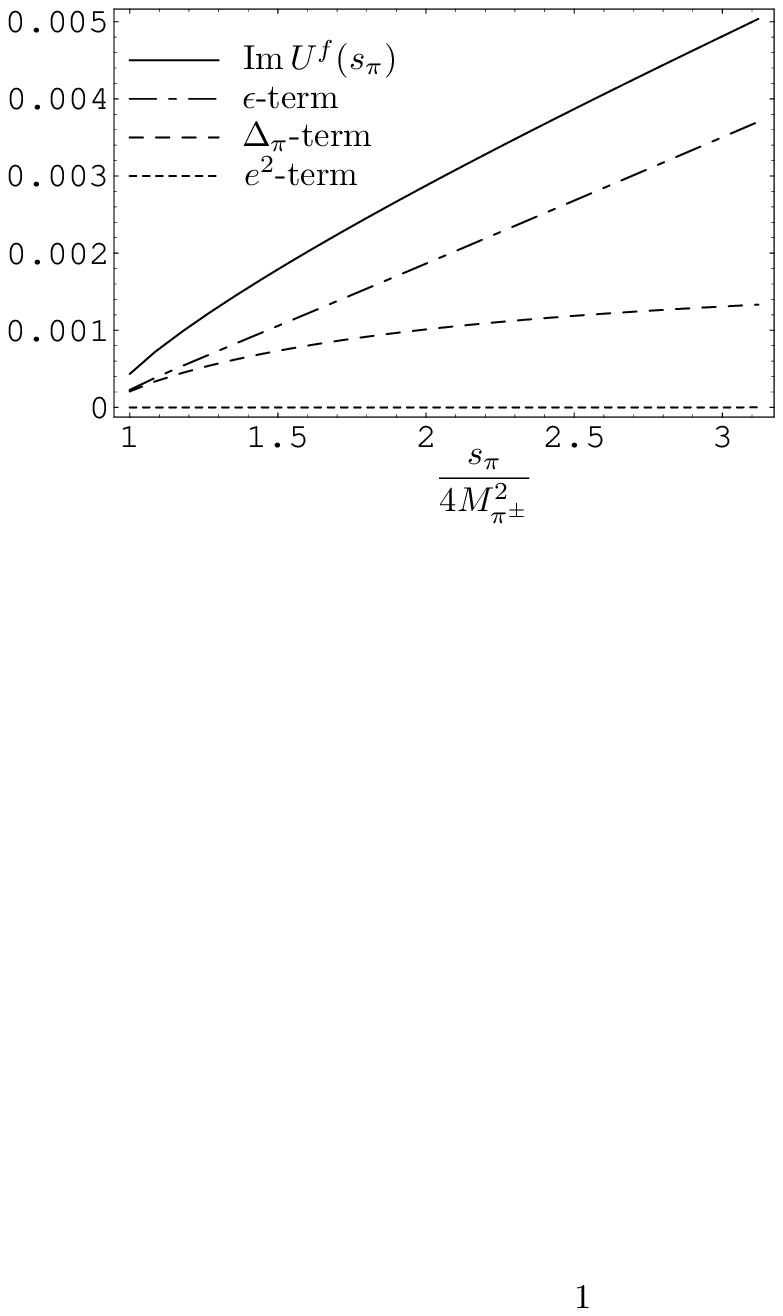}}
\caption{\label{fig:f-phase_electron} The imaginary part (in Radian) of the first term in the partial wave expansion for $f$ form factor under the assumptions, $s_{\ell}=m_{\ell}^2=m_{\mathrm{e}}^2$, $F_0=F_{\pi}=92.419$ MeV.}
\end{figure}

\begin{figure}[p]
\epsfxsize14cm \centerline{\epsffile{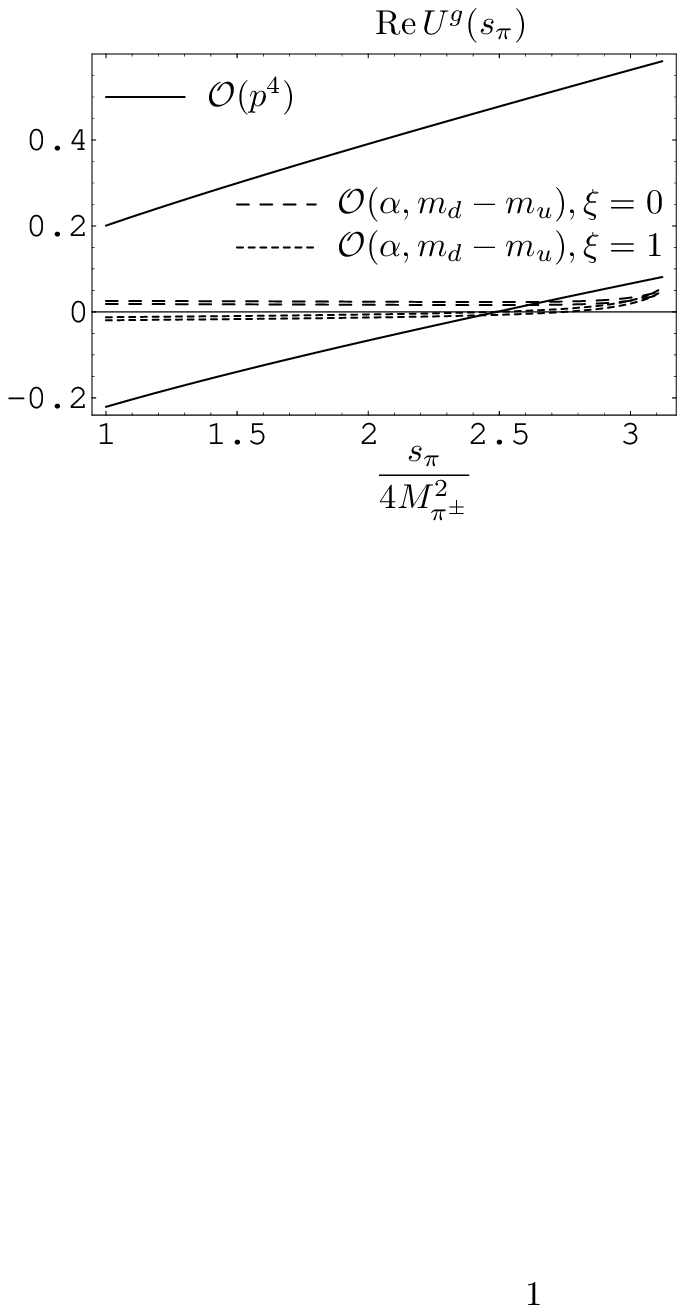}}
\caption{\label{fig:g-module_all} Radiative correction to the real
part of the first term in the partial wave expansion for $g$ form
factor under the assumptions,
$s_{\ell}=m_{\ell}^2=m_{\mathrm{e}}^2$, $F_0=67.53$ MeV. The
infrared divergence has been removed applying a minimal, $\xi =1$,
as well as a reasonable, $\xi =0$, subtraction scheme. Error bands
come exclusively from the uncertainty in the determination of
low-energy constants and have been developed in quadrature.}
\end{figure}

\begin{figure}[p]
\epsfxsize14cm \centerline{\epsffile{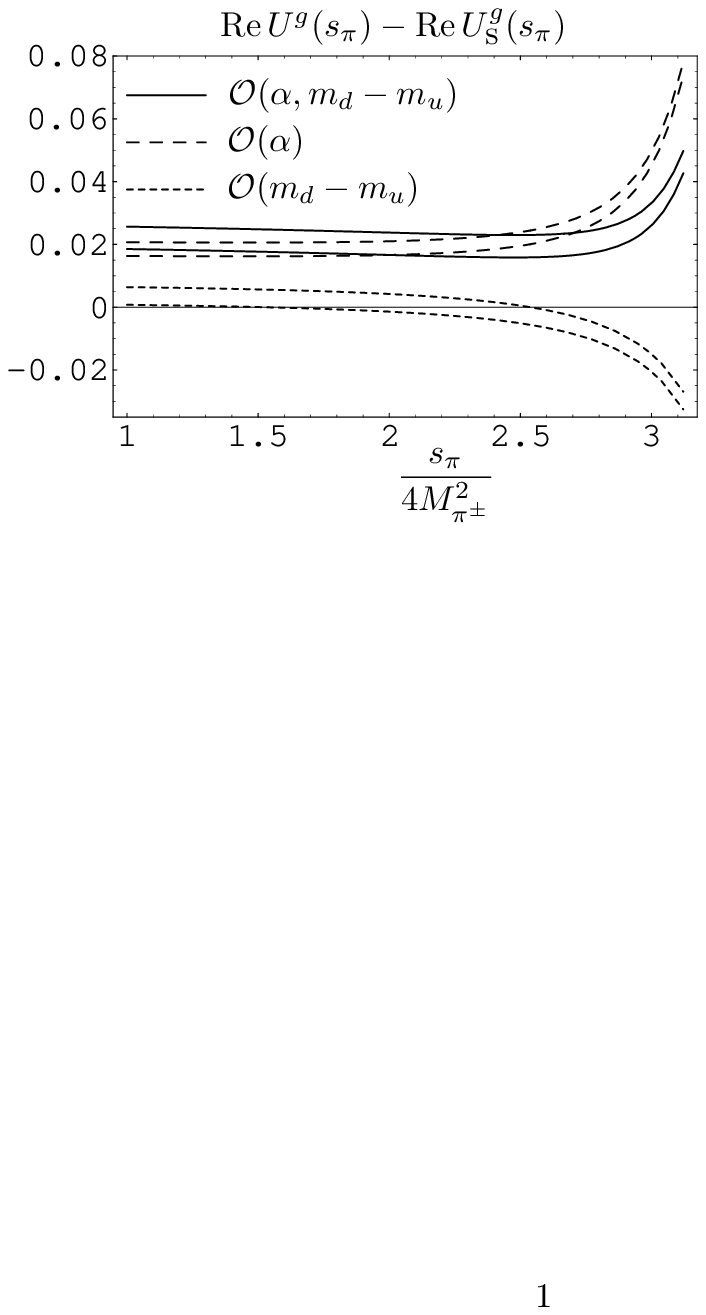}}
\caption{\label{fig:g-module_breaking} Isospin breaking correction
to the real part of the first term in the partial wave expansion
for $g$ form factor under the assumptions,
$s_{\ell}=m_{\ell}^2=m_{\mathrm{e}}^2$, $F_0=67.53$ MeV. The
infrared divergence has been removed applying a reasonable, $\xi
=0$, subtraction scheme. Error bands come exclusively from the
uncertainty in the determination of low-energy constants and have
been developed in quadrature.}
\end{figure}

\begin{figure}[p]
\epsfxsize14cm \centerline{\epsffile{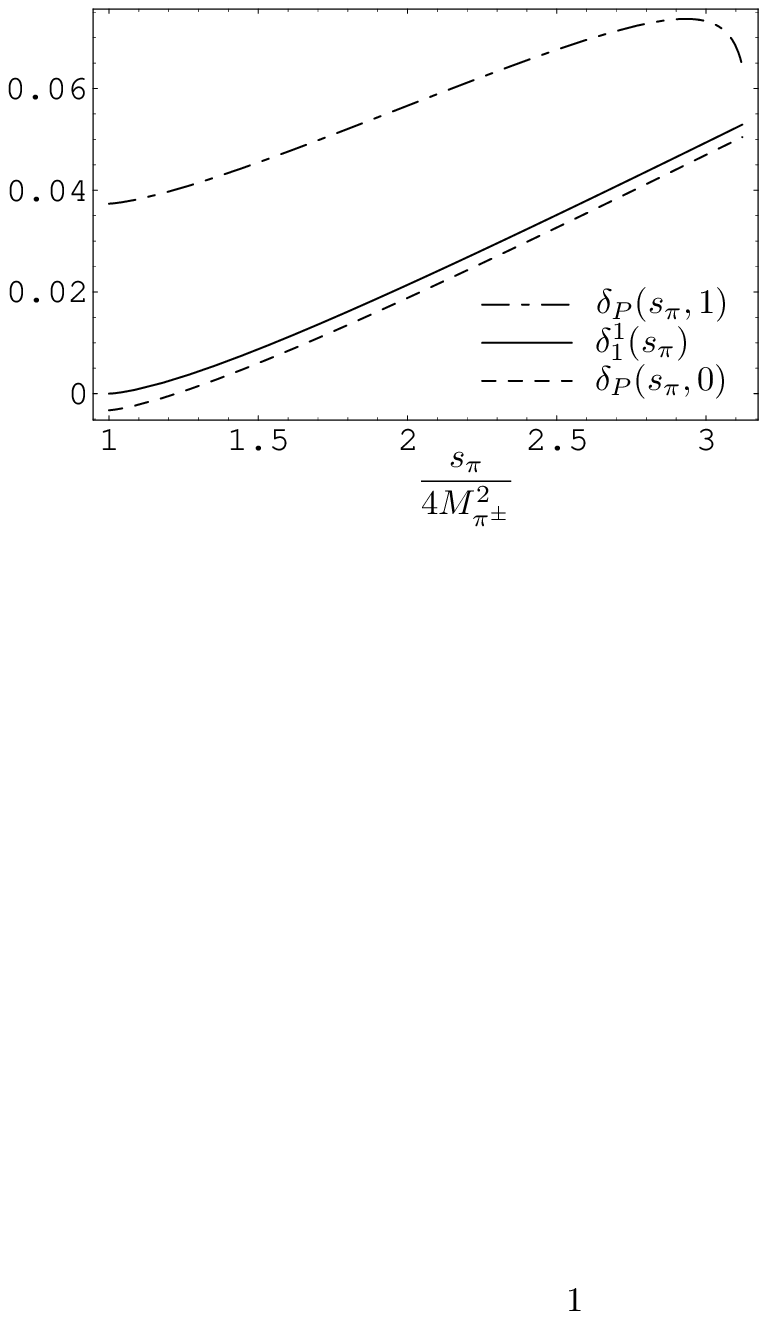}}
\caption{\label{fig:g-phase_electron} The imaginary part (in Radian) of the first term in the partial wave expansion for $g$ form factor under the assumptions, $s_{\ell}=m_{\ell}^2=m_{\mathrm{e}}^2$, $F_0=F_{\pi}=92.419$ MeV. The infrared divergence has been removed applying a minimal, $\xi =1$, as well as a reasonable, $\xi =0$, subtraction scheme.}
\end{figure}

\end{document}

%% file: introduction.tex
\section{Introduction}

Every time that a kaon decays into a couple of pions and a lepton-neutrino pair, a $\pi\pi$ scattering occurs in the final state. Whenever a pion scatters on its twin, it offers to us an additional opportunity to scrutinize the fundamental state of strong interaction (see~\cite{Nehme:2003bz} for references). Let $\delta_l^I$ be the phase of a two-pion state of angular momentum $l$ and Isospin $I$ and consider the $K_{\ell 4}$ decay process,
\begin{equation}
\label{eq:main_process} 
K(p) \longrightarrow \pi (p_1)\pi (p_2)\ell^+(p_{\ell})\nu_{\ell}(p_{\nu})\,,
\end{equation} 
where the lepton, $\ell$, is either a muon, $\mu$, or an electron, e, and $\nu$ stands for the corresponding neutrino. In the Isospin limit, the decay amplitude, $\mathcal{A}$, for process (\ref{eq:main_process}) can be parameterized in terms of three vectorial ($F$, $G$, and $R$) and one anomalous ($H$) form factors,
\begin{eqnarray}
\mathcal{A}
&\doteq& i\,\frac{G_F}{\sqrt{2}}\,V_{us}^*{\overline u}(\boldsymbol{p}_{\nu})\gamma_{\mu}(1-\gamma^5)v(\boldsymbol{p}_{\ell})\times
\nonumber \\ 
&& \left\lbrace \dfrac{i}{M_{K^{\pm}}}\left[ (p_1+p_2)^{\mu}F+(p_1-p_2)^{\mu}G+(p_{\ell}+p_{\nu})^{\mu}R\right] \right. 
\nonumber \\ 
&& \left. -\dfrac{1}{M_{K^{\pm}}^3}\,
\epsilon^{\mu\nu\rho\sigma}(p_{\ell}+p_{\nu})_{\nu}
(p_1+p_2)_{\rho}(p_1-p_2)_{\sigma}H\right\rbrace \,,
\end{eqnarray} 
where $V_{us}$ denotes the Cabibbo-Kobayashi-Maskawa flavor-mixing matrix element and $G_F$ is the so-called Fermi coupling constant. Note that form factors are made dimensionless by inserting the normalizations, $M_{K^{\pm}}^{-1}$ and $M_{K^{\pm}}^{-3}$. The fact that we have used the \textit{charged} kaon mass is a purely conventional matter and corresponds to the choice of defining the Isospin limit in terms of charged masses. In the following, we will be interested only in two form factors, $F$ and $G$, and denote by $(F,G)^{+-}$ and $(F,G)^{0-}$ the ones corresponding to the physical processes,
\begin{equation}
\label{eq:charged_process}
K^+(p)\longrightarrow\pi^+(p_1)\pi^-(p_2)\ell^+(p_{\ell})\nu_{\ell}(p_{\nu}), 
\end{equation} 
and
\begin{equation}
\label{eq:process}
K^0(p)\longrightarrow\pi^0(p_1)\pi^-(p_2)\ell^+(p_{\ell})\nu_{\ell}(p_{\nu}), 
\end{equation}
respectively. 

Form factors are analytic functions of three independent Lorentz invariants,
\begin{equation}
s_{\pi}\,\doteq\,(p_1+p_2)^2\,, \qquad s_{\ell}\,\doteq\,(p_{\ell}+p_{\nu})^2\,,
\end{equation} 
and the angle $\theta_{\pi}$ formed by $\boldsymbol{p}_1$, in the dipion rest frame, and the line of flight of the dipion as defined in the kaon rest frame~\cite{Cabibbo:1965, Cabibbo:1965E}. It has been shown in~\cite{Amoros:1999mg} that, in the experimentally relevant region, the partial wave expansion,
\begin{eqnarray}
F^{+-}
&=& \left( f_S(s_{\pi})+f_{\ell}\,s_{\ell}\right) e^{i\delta_0^0(s_{\pi})}
+\tilde{f}_PXY\cos\theta_{\pi}e^{i\delta_1^1(s_{\pi})}\,,
\\
G^{+-}
&=& \left( g_P+g_P's_{\pi}+g_{\ell}\,s_{\ell}\right) e^{i\delta_1^1(s_{\pi})}
+\tilde{g}_DXY\cos\theta_{\pi}e^{i\delta_2^0(s_{\pi})}\,,
\end{eqnarray}
is proving sufficient to parameterize form factors. In the preceding,
\begin{equation}
X\,\doteq\,\frac{1}{2}\,\lambda^{1/2}(s_{\pi},s_{\ell},M_{K^{\pm}}^2)\,, \quad 
Y\,\doteq\,\frac{1}{s_{\pi}}\,\lambda^{1/2}(s_{\pi},M_{\pi^{\pm}}^2,M_{\pi^{\pm}}^2)\,, 
\end{equation} 
with,
\begin{equation}
\lambda(x,y,z)\,\doteq\,x^2+y^2+z^2-2xy-2xz-2yz\,,
\end{equation} 
the usual K\"all\'en function. Note the linear dependence of the first term in the partial wave expansion of form factors on $s_{\ell}$. Isospin symmetry, Bose symmetry, and the $\Delta I=1/2$ rule lead to,
\begin{eqnarray}
F^{0-}
&=& \sqrt{2}\tilde{f}_PXY\cos\theta_{\pi}e^{i\delta_1^1(s_{\pi})}\,,
\label{eq:parameterization_strong_one}
\\
G^{0-}
&=& \sqrt{2}\left( g_P+g_P's_{\pi}+g_{\ell}\,s_{\ell}\right) e^{i\delta_1^1(s_{\pi})}\,.
\label{eq:parameterization_strong_two}
\end{eqnarray}
It follows that $K_{\ell 4}$ decay of the neutral kaon is dominated by $P$ waves. Therefore, a precise measurement of form factors for the decay in question would allow an accurate determination of the $P$-wave iso-vector $\pi\pi$ phase shift. 

The currently running KTeV experiment~\cite{Santos:2003} aims at measuring form factors for $K_{\ell 4}$ decay of the neutral kaon with an accuracy $3$ times better than the one offered by previous measurement~\cite{Makoff:1993xb, Makoff:1993xbE}. The outgoing data on form factors contain, besides strong interaction contribution, a contribution coming from electroweak interaction. The latter breaks Isospin symmetry and is expected to be sizeable near $\pi\pi$ production threshold~\cite{Stern:private}. In order to extract $\pi\pi$ scattering parameters from the KTeV measurement, Isospin breaking correction to form factors should therefore be under control. In this direction, we recently published analytic expressions for $F^{0-}$ and $G^{0-}$ form factors calculated at one-loop level in the framework of chiral perturbation theory based on the effective Lagrangian including mesons, photons, and leptons~\cite{Nehme:2003bz}. In the present work, we will split analytically Isospin limit and Isospin breaking part in form factors allowing a first evaluation of Isospin breaking effects in $K_{\ell 4}$ decays.    

%% file: section_1.tex
\section{Kinematical variables}

In the following, we shall consider process (\ref{eq:process}) and use, unless mentioned, notations of reference~\cite{Nehme:2003bz}. In the presence of Isospin breaking, the decay amplitude for process (\ref{eq:process}) can be written as follows by Lorentz covariance,
\begin{eqnarray}
{\cal A}^{0-}
&\doteq & \frac{G_FV_{us}^*}{\sqrt{2}}\,{\overline u}(\boldsymbol{p}_{\nu})
        (1+\gamma^5)\times
\nonumber \\
&& \left\lbrace \dfrac{1}{M_{K^{\pm}}}\left[ (p_1+p_2)^{\mu}f^{0-}+(p_1-p_2)^{\mu}g^{0-}+(p_{\ell}+p_{\nu})^{\mu}r^{0-}\right] \gamma_{\mu} \right.
\nonumber \\
&& +\dfrac{i}{M_{K^{\pm}}^3}\,
\epsilon^{\mu\nu\rho\sigma}(p_{\ell}+p_{\nu})_{\nu}
(p_1+p_2)_{\rho}(p_1-p_2)_{\sigma}h^{0-}
\nonumber \\
&& \left. +\frac{1}{2M_{K^{\pm}}^2}\,[\gamma_{\mu},\gamma_{\nu}]\,p_1^{\mu}p_2^{\nu}\,T\right\rbrace v(\boldsymbol{p}_l)\,.
\nonumber
\end{eqnarray}
The quantities $f$, $g$, $r$, and $h$, will be called the
\textit{corrected} $K_{\ell 4}$ form factors since their Isospin
limits are nothing else than the $K_{\ell 4}$ form factors, $F$, $G$,
$R$, and $H$, respectively. The tensorial form factor $T$ is purely Isospin breaking and does not contribute to the mixed process at leading chiral order. The corrected form factors as well as the tensorial one are analytic functions of five independent Lorentz invariants, $s_{\pi}$, $s_{\ell}$, $\theta_{\pi}$, $\theta_{\ell}$, and $\phi$. $\theta_{\ell}$ is the angle formed by $\boldsymbol{p}_{\ell}$, in the dilepton rest frame, and the line of flight of the dilepton as defined in the kaon rest frame. $\phi$ is the angle between the normals to the planes defined in the kaon rest frame by the pion pair and the lepton pair, respectively. Let us denote by $\delta F$ and $\delta G$ the next-to-leading
order corrections to the $F^{0-}$ and $G^{0-}$ form factors,
respectively,
\begin{eqnarray}
f^{0-}
&=& \frac{M_{K^{\pm}}}{F_0}\,\bigg (\,0+\delta F\,\bigg )\,,
\nonumber \\
g^{0-}
&=& \frac{M_{K^{\pm}}}{F_0}\,\bigg (\,1+\delta G\,\bigg )\,.
\nonumber
\end{eqnarray}
The analytic expressions for $\delta F$ and $\delta G$ were given
in~\cite{Nehme:2003bz}. We shall distinguish between
\textit{photonic} and \textit{non photonic} contributions to
$\delta F$ and $\delta G$. The photonic contribution comes from
those Feynman diagrams with a virtual photon exchanged between two
meson legs or one meson leg and a pure strong vertex. Obviously,
this contribution is proportional to $e^2$, where $e$ is the
electric charge, and depends in general on the five independent
kinematical variables, $s_{\pi}$, $s_{\ell}$, $\theta_{\pi}$,
$\theta_{\ell}$, and $\phi$ through Lorentz invariants like
$(p_2+p_{\ell})^2$, say. The non photonic contribution comes from
diagrams having similar topology as the ones in the pure strong
theory with Isospin breaking allowed in propagators and vertices.
This contribution generates Isospin breaking terms proportional to
the rate of $SU(2)$ to $SU(3)$ breaking,
\begin{equation}
\epsilon\,\doteq\,\dfrac{\sqrt{3}}{4}\,\dfrac{m_d-m_u}{m_s-\hat{m}}\,, \qquad
\hat{m}\,\doteq\,\frac{1}{2}\,(m_u+m_d)\,,
\end{equation}
and to mass square difference between charged and neutral mesons,
\begin{eqnarray}
\Delta_{\pi}
&\doteq& M_{\pi^{\pm}}^2-M_{\pi^0}^2\,=\,2Z_0e^2F_0^2+\mathcal{O}(p^4)\,,
\\
\Delta_K
&\doteq& M_{K^{\pm}}^2-M_{K^0}^2\,=\,2Z_0e^2F_0^2-B_0(m_d-m_u)+\mathcal{O}(p^4)\,,
\end{eqnarray}
or equivalently, $(m_d-m_u)/(m_s-\hat{m})$, $Z_0e^2$, and $m_d-m_u$. The kinematical dependence is on three Lorentz invariants, $(p_1+p_2)^2$, $(p-p_1)^2$, and $(p-p_2)^2$ which represent respectively the dipion mass square, the exchange energy between the kaon and the neutral pion, and that between the kaon and the charged pion. In terms of independent kinematical variables, the preceding scalars are functions of $s_{\pi}$, $s_{\ell}$, and $\cos\theta_{\pi}$.

\subsection{The photonic contribution}

A generic term in the photonic contribution can be,
\begin{equation}
\label{eq:photonic_contribution}
\mathrm{photonic~contribution}\,=\,e^2\sum_i\xi_i\left( (p_2+p_{\ell})^2,\ldots \right) \,,
\end{equation}
where $\xi_i$ is an arbitrary loop integral function of $(p_2+p_{\ell})^2$. To the order we are working, that is, to leading order in Isospin breaking, the power counting scheme we use dictates the following on-shell conditions to be used in the argument of $\xi_i$,
\begin{equation}
\label{eq:on-shell_conditions}
p^2\,=\,M_K^2\,\doteq\,B_0(m_s+\hat{m})\,, \quad p_1^2\,=\,p_2^2\,=\,M_{\pi}^2\,\doteq\,2B_0\hat{m}\,.
\end{equation}
Therefore, $(p_2+p_{\ell})^2$ in (\ref{eq:photonic_contribution}) should be replaced by the following expression~\cite{Nehme:2003bz},
\begin{eqnarray}
&& M_{\pi}^2+m_{\ell}^2
\nonumber \\
&+& \dfrac{1}{4}\left( 1+\dfrac{m_{\ell}^2}{s_{\ell}}\right) (M_K^2-s_{\ell}-s_{\pi})
\nonumber \\
&-& \dfrac{1}{4}\left( 1+\dfrac{m_{\ell}^2}{s_{\ell}}\right) \left( 1-\dfrac{4M_{\pi}^2}{s_{\pi}}\right) ^{1/2}\lambda^{1/2}(s_{\pi},s_{\ell},M_K^2)\cos\theta_{\pi}
\nonumber \\
&+& \dfrac{1}{4}\left( 1-\dfrac{m_{\ell}^2}{s_{\ell}}\right) \lambda^{1/2}(s_{\pi},s_{\ell},M_K^2)\cos\theta_{\ell}
\nonumber \\
&-& \dfrac{1}{4}\left( 1-\dfrac{m_{\ell}^2}{s_{\ell}}\right) \left( 1-\dfrac{4M_{\pi}^2}{s_{\pi}}\right) ^{1/2}(M_K^2-s_{\ell}-s_{\pi})\cos\theta_{\pi}\cos\theta_{\ell}
\nonumber \\
&+& \dfrac{1}{2}\left( 1-\dfrac{m_{\ell}^2}{s_{\ell}}\right) \left( 1-\dfrac{4M_{\pi}^2}{s_{\pi}}\right) ^{1/2}(s_{\pi}s_{\ell})^{1/2}\sin\theta_{\pi}\sin\theta_{\ell}\cos\phi\,.
\nonumber
\end{eqnarray}
From the foregoing, it is clear that for $s_{\ell}=m_{\ell}^2$ the photonic contribution does not depend neither on $\theta_{\ell}$ nor on $\phi$. In order to reduce the complexity of the study and allow the treatment of photonic and non photonic contributions to $\mathcal{A}^{0-}$ on an equal footing, we will assume that,
\begin{equation}
\label{eq:simplified_condition}
s_{\ell}\,=\,m_{\ell}^2\,,
\end{equation}
and use for $(p_2+p_{\ell})^2$ in (\ref{eq:photonic_contribution}) the following expression,
\begin{eqnarray}
\label{eq:photonic_scalar}
(p_2+p_{\ell})^2
&=& \dfrac{1}{2}\,(M_K^2+2M_{\pi}^2+m_{\ell}^2-s_{\pi})
\nonumber \\
&-& \dfrac{1}{2}\left( 1-\dfrac{4M_{\pi}^2}{s_{\pi}}\right) ^{1/2}\lambda^{1/2}(s_{\pi},m_{\ell}^2,M_K^2)\cos\theta_{\pi}\,.
\end{eqnarray}
It follows that (\ref{eq:photonic_contribution}) can be written as,
\begin{equation}
\label{eq:photonic_expansion}
\mathrm{photonic~contribution}\,=\,e^2\,\varsigma (s_{\pi})+e^2\,\vartheta (s_{\pi})\cos\theta_{\pi} \,,
\end{equation}
where $\varsigma$ and $\vartheta$ are analytic functions of $s_{\pi}$.

\subsection{The non photonic contribution}

In order to split strong and electromagnetic terms in the non photonic contribution, one has to expand the exchange energies, $(p-p_1)^2$ and $(p-p_2)^2$, in powers of the fine structure constant, $\alpha$, and $m_d-m_u$. To this end, we shall first express these scalars in terms of $s_{\pi}$ and $\cos\theta_{\pi}$ for $s_{\ell}=m_{\ell}^2$ and in the presence of Isospin breaking. From~\cite{Nehme:2003bz},
\begin{eqnarray}
(p-p_1)^2
&=& M_{K^0}^2+M_{\pi^0}^2
\nonumber \\
&-& \dfrac{1}{2s_{\pi}}\left[ (M_{K^0}^2-m_{\ell}^2+s_{\pi})(s_{\pi}+M_{\pi^0}^2-M_{\pi^{\pm}}^2)\right.
\nonumber \\
&+& \left.\lambda^{1/2}(s_{\pi},m_{\ell}^2,M_{K^0}^2)
\lambda^{1/2}(s_{\pi},M_{\pi^0}^2,M_{\pi^{\pm}}^2)\cos\theta_{\pi}\right] \,,
\label{eq:exchange_one}
\\
(p-p_2)^2
&=& M_{K^0}^2+M_{\pi^{\pm}}^2
\nonumber \\
&-& \dfrac{1}{2s_{\pi}}\left[ (M_{K^0}^2-m_{\ell}^2+s_{\pi})(s_{\pi}-M_{\pi^0}^2+M_{\pi^{\pm}}^2)\right.
\nonumber \\
&-& \left.\lambda^{1/2}(s_{\pi},m_{\ell}^2,M_{K^0}^2)
\lambda^{1/2}(s_{\pi},M_{\pi^0}^2,M_{\pi^{\pm}}^2)\cos\theta_{\pi}\right] \,.
\label{eq:exchange_two}
\end{eqnarray}
Let us denote by $t_{\pi}$ and $u_{\pi}$ the Isospin limits of the preceding Lorentz scalars,
\begin{eqnarray}
t_{\pi}
&=& \dfrac{1}{2}\,(M_{K^{\pm}}^2+2M_{\pi^{\pm}}^2+m_{\ell}^2-s_{\pi})
\nonumber \\
&-& \dfrac{1}{2}\left( 1-\dfrac{4M_{\pi^{\pm}}^2}{s_{\pi}}\right) ^{1/2}\lambda^{1/2}(s_{\pi},m_{\ell}^2,M_{K^{\pm}}^2)\cos\theta_{\pi}\,,
\label{eq:tpi}
\\
u_{\pi}
&=& \dfrac{1}{2}\,(M_{K^{\pm}}^2+2M_{\pi^{\pm}}^2+m_{\ell}^2-s_{\pi})
\nonumber \\
&+& \dfrac{1}{2}\left( 1-\dfrac{4M_{\pi^{\pm}}^2}{s_{\pi}}\right) ^{1/2}\lambda^{1/2}(s_{\pi},m_{\ell}^2,M_{K^{\pm}}^2)\cos\theta_{\pi}\,.
\label{eq:upi}
\end{eqnarray}
For completeness, it is convenient to note the following
proposition,
\begin{equation}
\label{eq:proposition}
\cos\theta_{\pi}\,=\,0\,\Longrightarrow\,t_{\pi}\,=\,u_{\pi}\,=\,
\dfrac{1}{2}\,(M_{K^{\pm}}^2+2M_{\pi^{\pm}}^2+m_{\ell}^2-s_{\pi})\,.
\end{equation}
Using the replacements,
\begin{equation}
\label{eq:replacements}
M_{\pi^0}^2\,\longrightarrow\,M_{\pi^{\pm}}^2-\Delta_{\pi}\,, \qquad
M_{K^0}^2\,\longrightarrow\,M_{K^{\pm}}^2-\Delta_K\,,
\end{equation}
and expanding (\ref{eq:exchange_one}) and (\ref{eq:exchange_two}) to first order in $\Delta_{\pi}$ and $\Delta_K$, we obtain,
\begin{eqnarray}
(p-p_1)^2
&=& \dfrac{1}{2}\,(M_{K^{\pm}}^2+2M_{\pi^{\pm}}^2+m_{\ell}^2-s_{\pi})
\nonumber \\
&+& \dfrac{1}{2s_{\pi}}\,(M_K^2-m_{\ell}^2-s_{\pi})\Delta_{\pi}-\dfrac{1}{2}\,\Delta_K
\nonumber \\
&+& \left[ -\dfrac{1}{2}\left( 1-\dfrac{4M_{\pi^{\pm}}^2}{s_{\pi}}\right) ^{1/2}\lambda^{1/2}(s_{\pi},m_{\ell}^2,M_{K^{\pm}}^2)\right.
\nonumber \\
&& -\dfrac{1}{2s_{\pi}}\left( 1-\dfrac{4M_{\pi}^2}{s_{\pi}}\right) ^{-1/2}\lambda^{1/2}(s_{\pi},m_{\ell}^2,M_K^2)\Delta_{\pi}
\nonumber \\
&& +\dfrac{1}{2}\left( 1-\dfrac{4M_{\pi}^2}{s_{\pi}}\right) ^{1/2}\times
\nonumber \\
&& (M_K^2-m_{\ell}^2-s_{\pi})\lambda^{-1/2}(s_{\pi},m_{\ell}^2,M_K^2)\Delta_K\,\bigg ] \cos\theta_{\pi}\,,
\label{eq:key_one}
\\
(p-p_2)^2
&=& \dfrac{1}{2}\,(M_{K^{\pm}}^2+2M_{\pi^{\pm}}^2+m_{\ell}^2-s_{\pi})
\nonumber \\
&-& \dfrac{1}{2s_{\pi}}\,(M_K^2-m_{\ell}^2+s_{\pi})\Delta_{\pi}-\dfrac{1}{2}\,\Delta_K
\nonumber \\
&+& \left[ \dfrac{1}{2}\left( 1-\dfrac{4M_{\pi^{\pm}}^2}{s_{\pi}}\right) ^{1/2}\lambda^{1/2}(s_{\pi},m_{\ell}^2,M_{K^{\pm}}^2)\right.
\nonumber \\
&& +\dfrac{1}{2s_{\pi}}\left( 1-\dfrac{4M_{\pi}^2}{s_{\pi}}\right) ^{-1/2}\lambda^{1/2}(s_{\pi},m_{\ell}^2,M_K^2)\Delta_{\pi}
\nonumber \\
&& -\dfrac{1}{2}\left( 1-\dfrac{4M_{\pi}^2}{s_{\pi}}\right) ^{1/2}\times
\nonumber \\
&& (M_K^2-m_{\ell}^2-s_{\pi})\lambda^{-1/2}(s_{\pi},m_{\ell}^2,M_K^2)\Delta_K\,\bigg ] \cos\theta_{\pi}\,.
\label{eq:key_two}
\end{eqnarray}
Note that terms of order $\mathcal{O}(\Delta_{\pi}\Delta_K)$ are forbidden by our power counting scheme since they are first order in Isospin breaking. Although equations (\ref{eq:key_one}) and (\ref{eq:key_two}) are simple to derive, their utility is of great importance to the present study. In fact, the involved expansion could be generalized to any $K_{\ell 4}$ observable as we will see below.

\subsection{Splitting strong and electromagnetic interactions}

The first step in our program consists on injecting equations (\ref{eq:key_one}) and (\ref{eq:key_two}) in the non photonic contribution to the decay amplitude $\mathcal{A}^{0-}$. Then, we expand once more to first order in $\Delta_{\pi}$ and $\Delta_K$ dropping out terms of order $\mathcal{O}(\Delta_{\pi}\Delta_K)$. As a result, form factors for $K_{\ell 4}$ decay of the neutral kaon can be written in the following compact form which shows explicitly the splitting between strong and electromagnetic interactions,
\begin{eqnarray}
\label{eq:principal_one}
&& x^{0-}\left( s_{\pi},(p-p_1)^2,(p-p_2)^2,(p_2+p_{\ell})^2,\ldots \right) \,=\,
\nonumber \\
&& \qquad \dfrac{M_{K^{\pm}}}{F_0}\left[ \delta_{xg}+U^x(s_{\pi})+V^x(s_{\pi})\cos\theta_{\pi}\right] \,,
\;\; x\,=\,f\,,\; g\,,
\end{eqnarray}
where,
\begin{eqnarray}
\label{eq:principal_two}
W^x
&=& W_{\mathrm{s}}^x+W_{\pi}^x\Delta_{\pi}+W_K^x\Delta_K
\nonumber \\
& & \qquad +W_{e^2}^xe^2+W_{\epsilon}^x\,\dfrac{\epsilon}{\sqrt{3}}\,, \quad W\,=\,U\,,\; V\,,
\end{eqnarray}
are analytic functions of $s_{\pi}$.
If one makes the following substitutions,
\begin{eqnarray}
\Delta_{\pi}
&\longrightarrow & 2Z_0e^2F_0^2\,,
\label{eq:difference_pi}
\\
\Delta_K
&\longrightarrow & 2Z_0e^2F_0^2-\dfrac{4\epsilon}{\sqrt{3}}\,(M_K^2-M_{\pi}^2)\,,
\label{eq:difference_K}
\end{eqnarray}
then, equations (\ref{eq:principal_one}) and (\ref{eq:principal_two}) read,
\begin{eqnarray}
W^x
&=& W_{\mathrm{s}}^x+W_{\alpha}^xe^2+W_{m_d-m_u}^x\,\dfrac{\epsilon}{\sqrt{3}}\,,
\label{eq:main_one}
\\
W_{\alpha}^x
&=& W_{e^2}^x+2Z_0F_0^2(W_{\pi}^x+W_K^x)\,,
\label{eq:main_two}
\\
W_{m_d-m_u}^x
&=& W_{\epsilon}^x-4(M_K^2-M_{\pi}^2)W_K^x\,.
\label{eq:main_three}
\end{eqnarray}
The aim of the present work is to determine the $U$ functions corresponding to $f$ and $g$ form factors for $K_{\ell 4}$ decay of the neutral kaon.

%% file: section_2.tex
\section{The photonic contribution}

From now on, we will work under proposition (\ref{eq:proposition}) keeping in mind that, in the Isospin breaking contribution, the power counting dictates the following, 
\begin{equation}
\mathrm{Isospin}\,\mathrm{breaking}\,\longrightarrow\,t_{\pi}\,=\,u_{\pi}\,=\,
\dfrac{1}{2}\,(M_K^2+2M_{\pi}^2+m_{\ell}^2-s_{\pi})\,.
\end{equation}
Taking the photonic contribution from~\cite{Nehme:2003bz}, applying assumption (\ref{eq:simplified_condition}), and performing the preceding expansion, it is easy at a first sight to derive $U_{e^2}$. The problem is that, in practice, one encounters loop integrals with vanishing Gramm determinant when reducing vector and tensor integrals to scalar ones~\cite{Melrose:1965kb}. After a long and tedious calculation one obtains,    
\begin{eqnarray}
U_{e^2}^f
&=& \dfrac{1}{3}\left( -6K_3+3K_4+2K_5+2K_6-6X_1\right) 
\nonumber \\ 
&-& \dfrac{2}{3}\,\dfrac{M_{\pi}^2}{M_{\pi}^2-M_{\eta}^2}\left( 6K_3-3K_4-2K_5-2K_6+2K_9+2K_{10}\right) 
\nonumber \\ 
&+& B(M_{\pi}^2,0,M_{\pi}^2)\,\bigg\{\,1
\nonumber \\ 
&-& \dfrac{1}{4}\left[ (M_K^2-m_{\ell}^2-s_{\pi})^2-4m_{\ell}^2M_{\pi}^2\right] \lambda^{-1}(t_{\pi},m_{\ell}^2,M_{\pi}^2)
\nonumber \\ 
&+& \dfrac{3}{4}\,m_{\ell}^2M_{\pi}^2(s_{\pi}-4M_{\pi}^2)
(M_K^2-m_{\ell}^2-s_{\pi})\lambda^{-2}(t_{\pi},m_{\ell}^2,M_{\pi}^2)\,\bigg\}
\nonumber \\ 
&+& B(m_{\ell}^2,0,m_{\ell}^2)\,\bigg\{\,-\dfrac{3}{4}\,m_{\ell}^2(M_K^2-m_{\ell}^2-s_{\pi})
\lambda^{-1}(t_{\pi},m_{\ell}^2,M_{\pi}^2)\,\bigg\}
\nonumber \\ 
&+& B(0,m_{\ell}^2,M_K^2)\,\bigg\{\,\dfrac{m_{\ell}^2}{4t_{\pi}}
\nonumber \\ 
&+& \dfrac{m_{\ell}^2}{4t_{\pi}}\,(s_{\pi}-4M_{\pi}^2)(M_{\pi}^2-m_{\ell}^2+t_{\pi})
\lambda^{-1}(t_{\pi},m_{\ell}^2,M_{\pi}^2)\,\bigg\}
\nonumber \\ 
&+& B(m_{\ell}^2,0,M_K^2)\,\bigg\{\,2m_{\ell}^4M_{\pi}^2(s_{\pi}-4M_{\pi}^2)
\lambda^{-2}(t_{\pi},m_{\ell}^2,M_{\pi}^2)
\nonumber \\ 
&+& \dfrac{1}{16}\,m_{\ell}^2(s_{\pi}-4M_{\pi}^2)(M_K^2-m_{\ell}^2-s_{\pi})^2
\lambda^{-2}(t_{\pi},m_{\ell}^2,M_{\pi}^2)\,\bigg\}
\nonumber \\ 
&+& B(t_{\pi},m_{\ell}^2,M_{\pi}^2)\,\bigg\{\,-1
+\dfrac{1}{2}\left[t_{\pi}(M_K^2-2m_{\ell}^2-s_{\pi})\right.
\nonumber \\ 
&-& \left.(M_{\pi}^2-m_{\ell}^2)(M_K^2-s_{\pi})\right] \lambda^{-1}(t_{\pi},m_{\ell}^2,M_{\pi}^2)\,\bigg\} 
\nonumber \\ 
&+& B(t_{\pi},M_{\pi}^2,M_K^2)\,\bigg\{\,-\dfrac{m_{\ell}^2}{4t_{\pi}}
\nonumber \\ 
&-& \dfrac{m_{\ell}^2}{4t_{\pi}}\,(s_{\pi}-4M_{\pi}^2)(M_{\pi}^2-m_{\ell}^2+t_{\pi})
\lambda^{-1}(t_{\pi},m_{\ell}^2,M_{\pi}^2)
\nonumber \\ 
&+& \dfrac{1}{4}\,m_{\ell}^2(s_{\pi}-4M_{\pi}^2)
\left[-t_{\pi}(3M_{\pi}^2-m_{\ell}^2+t_{\pi})\right. 
\nonumber \\ 
&+& \left.(M_{\pi}^2-m_{\ell}^2)(5M_{\pi}^2+m_{\ell}^2-t_{\pi})\right] \lambda^{-2}(t_{\pi},m_{\ell}^2,M_{\pi}^2)\,\bigg\} 
\nonumber \\ 
&+& C(M_{\pi}^2,t_{\pi},m_{\ell}^2,0,M_{\pi}^2,M_K^2)\times
\nonumber \\ 
&& \bigg\{\,m_{\ell}^2M_{\pi}^2(M_K^2-m_{\ell}^2)\lambda^{-1}(t_{\pi},m_{\ell}^2,M_{\pi}^2)
\nonumber \\ 
&+& \dfrac{3}{4}\,m_{\ell}^2M_{\pi}^2(s_{\pi}-4M_{\pi}^2)
(M_K^2-m_{\ell}^2)(M_K^2-m_{\ell}^2-s_{\pi})\lambda^{-2}(t_{\pi},m_{\ell}^2,M_{\pi}^2)\,\bigg\} 
\nonumber \\ 
&+& C(m_{\ell}^2,0,m_{\ell}^2,0,m_{\ell}^2,M_K^2)\times 
\nonumber \\ 
&& \bigg\{\,-\dfrac{1}{2}\,m_{\ell}^2(M_K^2-m_{\ell}^2)(M_{\pi}^2+m_{\ell}^2-t_{\pi})
\lambda^{-1}(t_{\pi},m_{\ell}^2,M_{\pi}^2)\,\bigg\}
\nonumber \\  
&+& C(t_{\pi},t_{\pi},0,m_{\ell}^2,M_{\pi}^2,M_K^2)\times 
\nonumber \\ 
&& \bigg\{\,-\dfrac{1}{2}\,m_{\ell}^2(M_K^2-m_{\ell}^2)(M_{\pi}^2-m_{\ell}^2+t_{\pi})
\lambda^{-1}(t_{\pi},m_{\ell}^2,M_{\pi}^2)
\nonumber \\ 
&+& \dfrac{m_{\ell}^2}{8t_{\pi}}\,(s_{\pi}-4M_{\pi}^2)(5M_{\pi}^2+m_{\ell}^2-t_{\pi})
(M_{\pi}^2-m_{\ell}^2-t_{\pi})\lambda^{-1}(t_{\pi},m_{\ell}^2,M_{\pi}^2)
\nonumber \\ 
&+& \dfrac{m_{\ell}^2}{16t_{\pi}}\,(s_{\pi}-4M_{\pi}^2)(M_K^2-m_{\ell}^2-s_{\pi})
(3M_{\pi}^2-3m_{\ell}^2-t_{\pi})\lambda^{-1}(t_{\pi},m_{\ell}^2,M_{\pi}^2)
\nonumber \\ 
&+& \dfrac{m_{\ell}^2}{4t_{\pi}}\,(M_{\pi}^2-m_{\ell}^2+t_{\pi})\,\bigg\}\,,
\label{eq:virtual_f}  \\ 
U_{e^2}^g
&=& -\dfrac{1}{18}\left( 24K_1+24K_2+8K_5+8K_6-36K_{12}+12X_1+9X_6\right) 
\nonumber \\ 
&+& \frac{1}{M_{\pi}^2}\,A(M_{\pi}^2)-\frac{1}{2m_{\ell}^2}\,A(m_{\ell}^2)
-\frac{1}{32\pi^2}\left( 5+2\ln\frac{m_{\gamma}^2}{M_{\pi}^2}+2\ln\frac{m_{\gamma}^2}{m_{\ell}^2}\right) 
\nonumber \\ 
&+& B(M_{\pi}^2,0,M_{\pi}^2)\,\bigg\{\,-m_{\ell}^2M_{\pi}^2
\lambda^{-1}(t_{\pi},m_{\ell}^2,M_{\pi}^2)
\nonumber \\ 
&-& (M_{\pi}^2+m_{\ell}^2-t_{\pi})(M_{\pi}^2-m_{\ell}^2+t_{\pi}) \lambda^{-1}(t_{\pi},m_{\ell}^2,M_{\pi}^2)
\nonumber \\ 
&-& \dfrac{3}{4}\,m_{\ell}^2M_{\pi}^2(s_{\pi}-4M_{\pi}^2)
(M_K^2-m_{\ell}^2-s_{\pi})\lambda^{-2}(t_{\pi},m_{\ell}^2,M_{\pi}^2)\,\bigg\}
\nonumber \\
&+& B(m_{\ell}^2,0,m_{\ell}^2)\,\bigg\{\,1
\nonumber \\ 
&+& \dfrac{1}{4}\,(M_K^2-m_{\ell}^2-s_{\pi})(M_K^2+2m_{\ell}^2-s_{\pi})
\lambda^{-1}(t_{\pi},m_{\ell}^2,M_{\pi}^2)\,\bigg\}
\nonumber \\ 
&+& B(0,m_{\ell}^2,M_K^2)\,\bigg\{\,\dfrac{1}{2}-\dfrac{m_{\ell}^2}{4t_{\pi}}
\nonumber \\ 
&-& \dfrac{m_{\ell}^2}{4t_{\pi}}\,(s_{\pi}-4M_{\pi}^2)(M_{\pi}^2-m_{\ell}^2+t_{\pi})
\lambda^{-1}(t_{\pi},m_{\ell}^2,M_{\pi}^2)\,\bigg\}
\nonumber \\ 
&+& B(m_{\ell}^2,0,M_K^2)\,\bigg\{\,-\dfrac{1}{2}-2m_{\ell}^4M_{\pi}^2(s_{\pi}-4M_{\pi}^2)
\lambda^{-2}(t_{\pi},m_{\ell}^2,M_{\pi}^2)
\nonumber \\ 
&-& \dfrac{1}{16}\,m_{\ell}^2(s_{\pi}-4M_{\pi}^2)(M_K^2-m_{\ell}^2-s_{\pi})^2
\lambda^{-2}(t_{\pi},m_{\ell}^2,M_{\pi}^2)\,\bigg\}
\nonumber \\ 
&+& B(t_{\pi},m_{\ell}^2,M_{\pi}^2)\,\bigg\{\,\dfrac{1}{2}\,
m_{\ell}^2(M_{\pi}^2-m_{\ell}^2)\lambda^{-1}(t_{\pi},m_{\ell}^2,M_{\pi}^2)
\nonumber \\
&-& \dfrac{1}{2}\,t_{\pi}(2M_K^2-3m_{\ell}^2-2s_{\pi}) \lambda^{-1}(t_{\pi},m_{\ell}^2,M_{\pi}^2)\,\bigg\} 
\nonumber \\ 
&+& B(t_{\pi},M_{\pi}^2,M_K^2)\,\bigg\{\,\dfrac{m_{\ell}^2}{4t_{\pi}}
\nonumber \\ 
&+& \dfrac{m_{\ell}^2}{4t_{\pi}}\,(s_{\pi}-4M_{\pi}^2)(M_{\pi}^2-m_{\ell}^2+t_{\pi})
\lambda^{-1}(t_{\pi},m_{\ell}^2,M_{\pi}^2)
\nonumber \\ 
&+& \dfrac{1}{4}\,m_{\ell}^2(s_{\pi}-4M_{\pi}^2)
\left[t_{\pi}(3M_{\pi}^2-m_{\ell}^2+t_{\pi})\right. 
\nonumber \\ 
&-& \left.(M_{\pi}^2-m_{\ell}^2)(5M_{\pi}^2+m_{\ell}^2-t_{\pi})\right] \lambda^{-2}(t_{\pi},m_{\ell}^2,M_{\pi}^2)\,\bigg\} 
\nonumber \\ 
&+& C(M_{\pi}^2,t_{\pi},m_{\ell}^2,0,M_{\pi}^2,M_K^2)\times
\nonumber \\ 
&& \bigg\{\,-m_{\ell}^2M_{\pi}^2(M_K^2-m_{\ell}^2)\lambda^{-1}(t_{\pi},m_{\ell}^2,M_{\pi}^2)
\nonumber \\ 
&-& \dfrac{3}{4}\,m_{\ell}^2M_{\pi}^2(s_{\pi}-4M_{\pi}^2)
(M_K^2-m_{\ell}^2)(M_K^2-m_{\ell}^2-s_{\pi})\lambda^{-2}(t_{\pi},m_{\ell}^2,M_{\pi}^2)\,\bigg\} 
\nonumber \\ 
&+& C(m_{\ell}^2,0,m_{\ell}^2,0,m_{\ell}^2,M_K^2)\times 
\nonumber \\ 
&& \bigg\{\,\dfrac{1}{2}\,m_{\ell}^2(M_K^2-m_{\ell}^2)(M_{\pi}^2+m_{\ell}^2-t_{\pi})
\lambda^{-1}(t_{\pi},m_{\ell}^2,M_{\pi}^2)\,\bigg\}
\nonumber \\  
&+& C(t_{\pi},t_{\pi},0,m_{\ell}^2,M_{\pi}^2,M_K^2)\times 
\nonumber \\ 
&& \bigg\{\,-\dfrac{m_{\ell}^2}{4t_{\pi}}\,(M_{\pi}^2-m_{\ell}^2+t_{\pi})
\nonumber \\ 
&+& \dfrac{1}{2}\,m_{\ell}^2(M_K^2-m_{\ell}^2)(M_{\pi}^2-m_{\ell}^2+t_{\pi})
\lambda^{-1}(t_{\pi},m_{\ell}^2,M_{\pi}^2)
\nonumber \\
&-& \dfrac{m_{\ell}^2}{8t_{\pi}}\,(s_{\pi}-4M_{\pi}^2)(5M_{\pi}^2+m_{\ell}^2-t_{\pi})
(M_{\pi}^2-m_{\ell}^2-t_{\pi})\lambda^{-1}(t_{\pi},m_{\ell}^2,M_{\pi}^2)
\nonumber \\ 
&+& \dfrac{1}{4}\,m_{\ell}^2(s_{\pi}-4M_{\pi}^2)(M_K^2-m_{\ell}^2-s_{\pi})
\lambda^{-1}(t_{\pi},m_{\ell}^2,M_{\pi}^2)
\nonumber \\
&-& \dfrac{3m_{\ell}^2}{16t_{\pi}}\,(s_{\pi}-4M_{\pi}^2)(M_K^2-m_{\ell}^2-s_{\pi})
(M_{\pi}^2-m_{\ell}^2+t_{\pi})\lambda^{-1}(t_{\pi},m_{\ell}^2,M_{\pi}^2)\,\bigg\}
\nonumber \\ 
&-& (M_K^2-m_{\ell}^2-s_{\pi})C(m_{\ell}^2,t_{\pi},M_{\pi}^2,m_{\gamma}^2,m_{\ell}^2,M_{\pi}^2)\,.
\label{eq:virtual_g}
\end{eqnarray} 

%% file: section_3.tex
\section{The non photonic contribution}

\subsection{One-point functions}

Let $P$ denotes a pion, $\pi$, or a kaon, $K$, and $\Delta_P$ the difference,
\begin{equation}
\label{eq:difference} 
\Delta_P\,\doteq\,M_{P^{\pm}}^2-M_{P^0}^2\,.
\end{equation} 
We shall expand the one-point function,
\begin{equation}
A(M_{P^0}^2)\,\doteq\,-i\mu^{4-D}\int\dfrac{d^Dl}{(2\pi )^D}\,\dfrac{1}{l^2-M_{P^0}^2}\,,
\end{equation} 
to leading order in Isospin breaking. 

In Dimensional Regularization, the preceding integral reads,
$$
A(M_{P^0}^2)\,=\,M_{P^0}^2\left[ -2\overline{\lambda}-\dfrac{1}{16\pi^2}\,\ln\left( \dfrac{M_{P^0}^2}{\mu^2}\right) \right] \,.
$$
By (\ref{eq:difference}), this is equivalent to,
\begin{eqnarray}
A(M_{P^0}^2)
&=& -2\overline{\lambda}(M_{P^{\pm}}^2-\Delta_P)
\nonumber \\ 
&-& \dfrac{1}{16\pi^2}\,(M_{P^{\pm}}^2-\Delta_P)\ln\left[ \left( \dfrac{M_{P^{\pm}}^2}{\mu^2}\right) \left( 1-\dfrac{\Delta_P}{M_P^2}\right) \right] \,.
\nonumber 
\end{eqnarray} 
Expanding to first order in $\Delta_P$, we obtain the splitting between strong and electromagnetic interactions in one-point functions,
\begin{equation}
\label{eq:splitting_one}
A(M_{P^0}^2)\,=\,A(M_{P^{\pm}}^2)+\left[ \dfrac{1}{16\pi^2}-\dfrac{1}{M_P^2}\,A(M_P^2)\right] \Delta_P\,. 
\end{equation} 

\subsection{Two-point functions}

The loop integral,
\begin{equation}
\label{eq:two-point_integral} 
B(p_1,m_0,m_1)\,\doteq\,-i\mu^{4-D}\int\dfrac{d^Dl}{(2\pi )^D}\,\dfrac{1}{(l^2-m_0^2)[(p_1+l)^2-m_1^2]}\,,
\end{equation} 
is function of three scalars, $p_1^2$, $m_0^2$, and $m_1^2$. In order to obtain Isospin breaking corrections generated from (\ref{eq:two-point_integral}) we shall expand $B(p_1^2+\delta ,m_0^2+\delta_0,m_1^2+\delta_1)$ to first order in $\delta$, $\delta_0$, and $\delta_1$, where these quantities are leading order in Isospin breaking,
\begin{equation}
\delta\,,\,\delta_0\,,\,\delta_1\,,\,=\,\mathcal{O}(\alpha\,,\,m_d-m_u)\,.
\end{equation}  
In Dimensional Regularization,
\begin{eqnarray}
&& B(p_1^2,m_0^2,m_1^2)\,=\,\dfrac{1}{16\pi^2}\left[ \dfrac{2}{4-D}+\ln (4\pi\mu^2)+\Gamma '(1)\right] 
\nonumber \\ 
&& -\dfrac{1}{16\pi^2}\int_0^1dx\,\ln\left[ xm_0^2+(1-x)m_1^2-x(1-x)p_1^2\right] \,.
\nonumber 
\end{eqnarray}
One then has,
\begin{eqnarray}
&& B(p_1^2+\delta ,m_0^2+\delta_0,m_1^2+\delta_1)\,=\,\dfrac{1}{16\pi^2}\left[ \dfrac{2}{4-D}+\ln (4\pi\mu^2)+\Gamma '(1)\right] 
\nonumber \\ 
&& -\dfrac{1}{16\pi^2}\int_0^1dx\,\ln\left[ x(m_0^2+\delta_0)+(1-x)(m_1^2+\delta_1)-x(1-x)(p_1^2+\delta )\right] \,.
\nonumber 
\end{eqnarray} 
Expanding to first order in $\delta$, $\delta_0$, and $\delta_1$, the preceding equation takes the form,
\begin{eqnarray}
&& B(p_1^2+\delta ,m_0^2+\delta_0,m_1^2+\delta_1)\,=\,B(p_1^2,m_0^2,m_1^2)
\nonumber \\ 
&& -\dfrac{1}{16\pi^2}\int_0^1dx
\dfrac{x}{xm_0^2+(1-x)m_1^2-x(1-x)p_1^2}\,\cdot\,\delta_0
\nonumber \\ 
&& -\dfrac{1}{16\pi^2}\int_0^1dx
\dfrac{1-x}{xm_0^2+(1-x)m_1^2-x(1-x)p_1^2}\,\cdot\,\delta_1
\nonumber \\ 
&& -\dfrac{1}{16\pi^2}\int_0^1dx
\dfrac{-x(1-x)}{xm_0^2+(1-x)m_1^2-x(1-x)p_1^2}\,\cdot\,\delta\,.
\nonumber 
\end{eqnarray} 
If we denote by $\tau$ the generic integral,
\begin{equation}
\label{eq:tau_definition} 
\tau (p_1^2,m_0^2,m_1^2)\,\doteq\,
\int_0^1dx\dfrac{1}{xm_0^2+(1-x)m_1^2-x(1-x)p_1^2}\,,
\end{equation} 
then, the splitting between strong and electromagnetic interactions in two-point functions is easily obtained from the following compact formula,
\begin{eqnarray}
\label{eq:splitting_two}
&& B(p_1^2+\delta ,m_0^2+\delta_0,m_1^2+\delta_1)\,=\,B(p_1^2,m_0^2,m_1^2)
\nonumber \\ 
&& -\dfrac{1}{32\pi^2p_1^2}\left[ \ln\left( \dfrac{m_0^2}{m_1^2}\right) +(p_1^2+m_1^2-m_0^2)\tau (p_1^2,m_0^2,m_1^2)\right] \delta_0
\nonumber \\ 
&& +\dfrac{1}{32\pi^2p_1^2}\left[ \ln\left( \dfrac{m_0^2}{m_1^2}\right) -(p_1^2-m_1^2+m_0^2)\tau (p_1^2,m_0^2,m_1^2)\right] \delta_1
\nonumber \\  
&& -\dfrac{1}{32\pi^2p_1^4}\left\lbrace 2p_1^2+(m_1^2-m_0^2)\ln\left( \dfrac{m_0^2}{m_1^2}\right) \right. 
\nonumber \\ 
&& +\left[ (m_1^2-m_0^2)^2-p_1^2(m_1^2+m_0^2)\right] \tau (p_1^2,m_0^2,m_1^2)\,\bigg\}\,\delta\,.
\end{eqnarray} 

As an application, consider the two-point function, $B(p_1-p,M_{\pi^0},M_{K^0})$, selected from the $t$-channel contribution to $\mathcal{A}^{0-}$. The following replacements in (\ref{eq:splitting_two}),
$$
\delta\longrightarrow -\dfrac{1}{s_{\pi}}\,(M_{\pi}^2+m_{\ell}^2-t_{\pi})\Delta_{\pi}-\dfrac{1}{2}\,\Delta_K\,, 
$$
$$
\delta_0\longrightarrow-\Delta_{\pi}\,, \qquad \delta_1\longrightarrow -\Delta_K\,,
$$
lead to the expression,
\begin{eqnarray}
&& B(p_1-p,M_{\pi^0},M_{K^0})\,=\,B(t_{\pi},M_{\pi^{\pm}}^2,M_{K^{\pm}}^2)
\nonumber \\ 
&& +\dfrac{\Delta_{\pi}}{32\pi^2t_{\pi}}\left\lbrace \dfrac{2}{s_{\pi}}\,(M_{\pi}^2+m_{\ell}^2-t_{\pi})\right. 
\nonumber \\ 
&& +\left[ 1+\dfrac{1}{s_{\pi}t_{\pi}}\,(M_{\pi}^2+m_{\ell}^2-t_{\pi})(M_K^2-M_{\pi}^2)\right] \ln\left( \dfrac{M_{\pi}^2}{M_K^2}\right) 
\nonumber \\ 
&& +\left[ M_K^2-M_{\pi}^2+t_{\pi}-\dfrac{1}{s_{\pi}}\,(M_K^2+M_{\pi}^2)\right. 
\nonumber \\ 
&& \left.\left. +\dfrac{1}{s_{\pi}t_{\pi}}\,(M_{\pi}^2+m_{\ell}^2-t_{\pi})(M_K^2-M_{\pi}^2)^2\right] \tau (t_{\pi},M_{\pi}^2,M_K^2)\right\rbrace 
\nonumber \\ 
&& +\dfrac{\Delta_K}{32\pi^2t_{\pi}}\left\lbrace 1+\dfrac{1}{2t_{\pi}}\,(M_K^2-M_{\pi}^2-2t_{\pi})\ln\left( \dfrac{M_{\pi}^2}{M_K^2}\right) \right. 
\nonumber \\ 
&& -\left[ \dfrac{1}{2}\,(3M_K^2-M_{\pi}^2-2t_{\pi})\right. 
\nonumber \\ 
&& \left.\left. -\dfrac{1}{2t_{\pi}}\,(M_K^2-M_{\pi}^2)^2\right] \tau (t_{\pi},M_{\pi}^2,M_K^2)\right\rbrace \,.
\end{eqnarray}

\subsection{Isospin limit}

\begin{eqnarray}
U_{\mathrm{s}}^f
&=& 0\,, 
\\ 
U_{\mathrm{s}}^g
&=& -\dfrac{1}{24F_0^2}\,\bigg\{\,\dfrac{1}{16\pi^2}\,\bigg [\,4(2M_{K^{\pm}}^2+4M_{\pi^{\pm}}^2-s_{\pi}) 
\nonumber \\ 
& & +(2M_{K^{\pm}}^2+4M_{\pi^{\pm}}^2-t_{\pi})
-\dfrac{3}{t_{\pi}}\,(M_{\pi^{\pm}}^2-M_{K^{\pm}}^2)(M_{\pi^{\pm}}^2+M_{K^{\pm}}^2) 
\nonumber \\ 
& & +(6M_{K^{\pm}}^2-t_{\pi})+\dfrac{9}{t_{\pi}}\,(M_{\eta}^2-M_{K^{\pm}}^2)
(M_{\eta}^2+M_{K^{\pm}}^2)\,\bigg ] 
\nonumber \\ 
&+& 48\,\bigg [\,2(M_{\pi^{\pm}}^2+M_{K^{\pm}}^2-t_{\pi})L_3
+2(M_{\pi^{\pm}}^2+2M_{K^{\pm}}^2)L_4-m_{\ell}^2L_9\,\bigg ]
\nonumber \\ 
&+& A(M_{\pi^{\pm}}^2)\,\bigg [\,5-\dfrac{6}{t_{\pi}}\,M_{\pi^{\pm}}^2
-\dfrac{6}{t_{\pi}^2}\,(M_{\pi^{\pm}}^2-M_{K^{\pm}}^2)^2\,\bigg ]
\nonumber \\ 
&+& A(M_{\eta}^2)\,\bigg [\,-3-\dfrac{6}{t_{\pi}}\,(5M_{\eta}^2-6M_{K^{\pm}}^2)
+\dfrac{18}{t_{\pi}^2}\,(M_{\eta}^2-M_{K^{\pm}}^2)^2\,\bigg ]
\nonumber \\ 
&+& A(M_{K^{\pm}}^2)\,\bigg [\,-2+\dfrac{12}{t_{\pi}}\,(M_{\pi^{\pm}}^2-M_{K^{\pm}}^2)
\nonumber \\ 
& & +\dfrac{12}{t_{\pi}}\,(M_{\eta}^2-M_{K^{\pm}}^2)
+\dfrac{6}{t_{\pi}^2}\,(M_{\pi^{\pm}}^2-M_{K^{\pm}}^2)^2
-\dfrac{18}{t_{\pi}^2}\,(M_{\eta}^2-M_{K^{\pm}}^2)^2\,\bigg ]
\nonumber \\ 
&+& 4(4M_{\pi^{\pm}}^2-s_{\pi})B(s_{\pi},M_{\pi^{\pm}}^2,M_{\pi^{\pm}}^2)
\nonumber \\ 
&+& 2(4M_{K^{\pm}}^2-s_{\pi})B(s_{\pi},M_{K^{\pm}}^2,M_{K^{\pm}}^2)
\nonumber \\ 
&+& B(t_{\pi},M_{\pi^{\pm}}^2,M_{K^{\pm}}^2)\,\bigg [\,-6(M_{\pi^{\pm}}^2-M_{K^{\pm}}^2+t_{\pi})
\nonumber \\ 
& & +\dfrac{6}{t_{\pi}}\,(M_{\pi^{\pm}}^2-M_{K^{\pm}}^2)^2
+\dfrac{6}{t_{\pi}^2}\,(M_{\pi^{\pm}}^2-M_{K^{\pm}}^2)^3\,\bigg ]
\nonumber \\ 
&+& B(t_{\pi},M_{\eta}^2,M_{K^{\pm}}^2)\,\bigg [\,-6(M_{\eta}^2-3M_{K^{\pm}}^2+t_{\pi})
\nonumber \\ 
& & +\dfrac{6}{t_{\pi}}\,(M_{\eta}^2-M_{K^{\pm}}^2)(5M_{\eta}^2-3M_{K^{\pm}}^2)
-\dfrac{18}{t_{\pi}^2}\,(M_{\eta}^2-M_{K^{\pm}}^2)^3\,\bigg ]\,\bigg\}\,. 
\end{eqnarray}

\subsection{The $\epsilon$-terms}

\begin{eqnarray}
U_{\epsilon}^f
&=& -3
\nonumber \\ 
&+& \dfrac{1}{24F_0^2}\,\dfrac{1}{16\pi^2}\,\bigg [\,2(6M_{\eta}^2+28M_K^2+20M_{\pi}^2-9t_{\pi})
\nonumber \\ 
&& +\dfrac{15}{t_{\pi}}\,(M_{\pi}^2-M_K^2)(M_{\pi}^2+M_K^2)
-\dfrac{9}{t_{\pi}}\,(M_{\eta}^2-M_K^2)(M_{\eta}^2+M_K^2)\,\bigg ]
\nonumber \\
&+& \dfrac{2}{F_0^2}\,\bigg [\,2(M_K^2+5M_{\pi}^2-2s_{\pi}-t_{\pi})L_3
\nonumber \\
&& +6(M_{\pi}^2+2M_K^2)L_4+48(M_K^2-M_{\pi}^2)(3L_7+L_8)-3m_{\ell}^2L_9\,\bigg ]
\nonumber \\ 
&+& \dfrac{1}{8F_0^2}\,\bigg\{\,-A(M_{\pi}^2)\,\bigg [\,\dfrac{32M_{\pi}^2}{M_{\pi}^2-M_{\eta}^2}
\nonumber \\ 
&& +15+\dfrac{2}{t_{\pi}}\,M_{\pi}^2
-\dfrac{10}{t_{\pi}^2}\,(M_{\pi}^2-M_K^2)^2\,\bigg ]
\nonumber \\ 
&+& 3A(M_{\eta}^2)\,\bigg [3+\dfrac{2}{t_{\pi}}\,(3M_{\eta}^2-4M_K^2)
-\dfrac{2}{t_{\pi}^2}\,(M_{\eta}^2-M_K^2)^2\,\bigg ]
\nonumber \\ 
&+& 2A(M_K^2)\,\bigg [\,\dfrac{16M_{\pi}^2}{M_{\pi}^2-M_{\eta}^2}
\nonumber \\ 
&& +3+\dfrac{1}{t_{\pi}}\,(5M_K^2-4M_{\pi}^2)
-\dfrac{5}{t_{\pi}^2}\,(M_{\pi}^2-M_K^2)^2
\nonumber \\ 
&& -\dfrac{3}{t_{\pi}}\,(2M_{\eta}^2-3M_K^2)
+\dfrac{3}{t_{\pi}^2}\,(M_{\eta}^2-M_K^2)^2\,\bigg ]
\nonumber \\ 
&-& 12(2M_{\pi}^2-s_{\pi})B(s_{\pi},M_{\pi}^2,M_{\pi}^2)
\nonumber \\ 
&+& 2(4M_K^2-3s_{\pi})B(s_{\pi},M_K^2,M_K^2)
\nonumber \\ 
&+& 4(3M_{\eta}^2+M_{\pi}^2-3s_{\pi})B(s_{\pi},M_{\eta}^2,M_{\pi}^2)
\nonumber \\ 
&+& 2B(t_{\pi},M_{\pi}^2,M_K^2)\,\bigg [\,7M_K^2+13M_{\pi}^2-9t_{\pi}
\nonumber \\ 
&& +\dfrac{1}{t_{\pi}}\,(M_{\pi}^2-M_K^2)(5M_K^2+M_{\pi}^2)
-\dfrac{5}{t_{\pi}^2}\,(M_{\pi}^2-M_K^2)^3\,\bigg ]
\nonumber \\ 
&+& 2B(t_{\pi},M_{\eta}^2,M_K^2)\,\bigg [\,3M_{\eta}^2-5M_K^2+3t_{\pi}
\nonumber \\ 
&& -\dfrac{9}{t_{\pi}}\,(M_{\eta}^2-M_K^2)^2
+\dfrac{3}{t_{\pi}^2}\,(M_{\eta}^2-M_K^2)^3\,\bigg ]\,\bigg\}\,, 
\\ 
U_{\epsilon}^g
&=& \dfrac{1}{4F_0^2}\,\bigg\{\,3\left( 1+\dfrac{2}{t_{\pi}}\,M_K^2\right) A(M_{\pi}^2)
\nonumber \\ 
&-& \left[ 3-\dfrac{2}{t_{\pi}}\,(3M_{\eta}^2-4M_K^2)\right] A(M_{\eta}^2)
\nonumber \\ 
&-& 2\left[ \dfrac{3}{t_{\pi}}\,M_K^2+\dfrac{1}{t_{\pi}}\,(3M_{\eta}^2-4M_K^2)\right] A(M_K^2)
\nonumber \\ 
&-& \dfrac{6}{t_{\pi}}\,M_K^2(M_{\pi}^2-M_K^2)B(t_{\pi},M_{\pi}^2,M_K^2)
\nonumber \\ 
&+& 2B(t_{\pi},M_{\eta}^2,M_K^2)\times
\nonumber \\ 
&& \left[ 3(M_{\eta}^2-M_K^2)
-\dfrac{1}{t_{\pi}}\,(M_{\eta}^2-M_K^2)(3M_{\eta}^2-4M_K^2)\right] \,\bigg\}\,.
\end{eqnarray} 

\subsection{The $\Delta_{\pi}$-terms}

\begin{eqnarray}
U_{\pi}^f
&=& -\dfrac{2}{F_0^2}\left( 1-\dfrac{\Delta_{\ell K}}{s_{\pi}}\right) L_3
\nonumber \\ 
&-& \dfrac{1}{192\pi^2F_0^2}\,\dfrac{1}{s_{\pi}}\,(4M_K^2+20M_{\pi}^2-3s_{\pi})
\nonumber \\ 
&-& \dfrac{1}{256\pi^2F_0^2}\,\dfrac{1}{s_{\pi}}\,\bigg\{\,4(M_{\pi}^2+m_{\ell}^2-t_{\pi})
\nonumber \\ 
&+& \dfrac{6}{t_{\pi}}\,(M_{\eta}^2+M_{\pi}^2)\Delta_{\ell K}
-\dfrac{1}{t_{\pi}}\,(M_{\eta}^2+6M_K^2+M_{\pi}^2)s_{\pi}
\nonumber \\ 
&+& \dfrac{1}{t_{\pi}^2}\left[ (5M_K^2-3M_{\pi}^2)\Delta_{\pi K}-5(3M_{\eta}^2-M_K^2)\Delta_{\eta K}\right] \Delta_{\ell K}
\nonumber \\ 
&-& \dfrac{2}{t_{\pi}^3}\,(\Delta_{\pi K}^2+\Delta_{\eta K}^2)\Delta_{\ell K}\Delta_{\pi K}\,\bigg\}
\nonumber \\ 
&+& \dfrac{1}{24F_0^2}\,\dfrac{1}{s_{\pi}}\,A(M_{\pi}^2)\,\bigg [\,40+\dfrac{6}{t_{\pi}}\,\dfrac{s_{\pi}}{M_{\pi}^2}\,\Delta_{\pi K}
\nonumber \\ 
&& +\dfrac{3}{t_{\pi}^2}\,\Delta_{\pi K}s_{\pi}+\dfrac{3}{t_{\pi}^2}\,(M_K^2-2M_{\pi}^2)\Delta_{\ell K}
-\dfrac{6}{t_{\pi}^3}\,\Delta_{\ell K}\Delta_{\pi K}^2\,\bigg ]
\nonumber \\ 
&-& \dfrac{1}{8F_0^2}\,\dfrac{1}{s_{\pi}}\,A(M_{\eta}^2)\,\bigg [\,\dfrac{1}{t_{\pi}}\,s_{\pi}-\dfrac{1}{t_{\pi}^2}\,\Delta_{\eta K}s_{\pi}
\nonumber \\ 
&& +\dfrac{1}{t_{\pi}^2}\,(6M_{\eta}^2-7M_K^2)\Delta_{\ell K}
+\dfrac{2}{t_{\pi}^3}\,\Delta_{\ell K}\Delta_{\pi K}\Delta_{\eta K}\,\bigg ]
\nonumber \\ 
&+& \dfrac{1}{24F_0^2}\,\dfrac{1}{s_{\pi}}\,A(M_K^2)\,\bigg [\,8-\dfrac{9}{t_{\pi}}\,s_{\pi}-\dfrac{3}{t_{\pi}^2}\,(\Delta_{\pi K}+\Delta_{\eta K})s_{\pi}
\nonumber \\ 
&& +\dfrac{9}{t_{\pi}^2}\,(M_{\eta}^2-2M_K^2+M_{\pi}^2)\Delta_{\ell K}
+\dfrac{6}{t_{\pi}^3}\,(\Delta_{\pi K}+\Delta_{\eta K})\Delta_{\ell K}\Delta_{\pi K}\,\bigg ]
\nonumber \\ 
&-& \dfrac{1}{256\pi^2F_0^2}\,\dfrac{1}{s_{\pi}}\,\ln\left( \dfrac{M_{\pi}^2}{M_K^2}\right) \,\bigg [\, \dfrac{2}{t_{\pi}}\,M_K^2s_{\pi}+\dfrac{2}{t_{\pi}^2}\,M_K^2\Delta_{\pi K}s_{\pi}
\nonumber \\ 
&& -\dfrac{1}{t_{\pi}^2}\,(M_K^2+3M_{\pi}^2)\Delta_{\ell K}\Delta_{\pi K}+\dfrac{2}{t_{\pi}^3}\,\Delta_{\ell K}\Delta_{\pi K}^3+\dfrac{1}{t_{\pi}^4}\,\Delta_{\ell K}\Delta_{\pi K}^4\,\bigg ]
\nonumber \\ 
&+& \dfrac{1}{512\pi^2F_0^2}\,\dfrac{1}{s_{\pi}}\,\ln\left( \dfrac{M_{\eta}^2}{M_K^2}\right) \,\bigg [\,\dfrac{2}{t_{\pi}^2}\,(3M_{\eta}^2-M_K^2)\Delta_{\ell K}\Delta_{\eta K}
\nonumber \\ 
&& -\dfrac{4}{t_{\pi}^3}\,(3M_{\eta}^2-2M_K^2)\Delta_{\ell K}\Delta_{\eta K}^2-\dfrac{2}{t_{\pi}^4}\,\Delta_{\ell K}\Delta_{\pi K}\Delta_{\eta K}^3\,\bigg ]
\nonumber \\ 
&-& \dfrac{1}{256\pi^2F_0^2}\,\dfrac{1}{s_{\pi}}\,\tau (t_{\pi},M_{\pi}^2,M_K^2)\times
\nonumber \\ 
&& \bigg [\,2M_K^2s_{\pi}-\dfrac{1}{t_{\pi}}\,(M_K^2+3M_{\pi}^2)\Sigma_{\pi K}\Delta_{\ell K}
\nonumber \\ 
&& +\dfrac{1}{t_{\pi}^2}\,(3M_K^2+5M_{\pi}^2)\Delta_{\ell K}\Delta_{\pi K}^2-\dfrac{2}{t_{\pi}^2}\,M_K^2\Delta_{\pi K}^2s_{\pi}
\nonumber \\ 
&& +\dfrac{1}{t_{\pi}^3}\,(3M_K^2-M_{\pi}^2)\Delta_{\ell K}\Delta_{\pi K}^3-\dfrac{1}{t_{\pi}^4}\,\Delta_{\ell K}\Delta_{\pi K}^5\,\bigg ]
\nonumber \\ 
&-& \dfrac{1}{512\pi^2F_0^2}\,\dfrac{1}{s_{\pi}}\,\tau (t_{\pi},M_{\eta}^2,M_K^2)\Delta_{\ell K}(\Delta_{\eta K}^2-\Sigma_{\eta K}t_{\pi})\times
\nonumber \\ 
&& \bigg [\,\dfrac{2}{t_{\pi}^2}\,(3M_{\eta}^2-M_K^2)
-\dfrac{4}{t_{\pi}^3}\,(3M_{\eta}^2-2M_K^2)\Delta_{\eta K}
-\dfrac{2}{t_{\pi}^4}\,\Delta_{\pi K}\Delta_{\eta K}^2\,\bigg ]
\nonumber \\ 
&-& \dfrac{1}{3F_0^2}\,\dfrac{1}{s_{\pi}}\,(5M_{\pi}^2-2s_{\pi})
B(s_{\pi},M_{\pi}^2,M_{\pi}^2)
\nonumber \\ 
&-& \dfrac{1}{3F_0^2}\,\dfrac{1}{s_{\pi}}\,(M_K^2-s_{\pi})B(s_{\pi},M_K^2,M_K^2)
\nonumber \\ 
&-& \dfrac{1}{8F_0^2}\,\dfrac{1}{s_{\pi}}\,B(t_{\pi},M_{\pi}^2,M_K^2)\,\bigg [\,s_{\pi}-\dfrac{1}{t_{\pi}}\,(5M_K^2-2M_{\pi}^2)s_{\pi}
\nonumber \\ 
&& +\dfrac{1}{t_{\pi}^2}\,(2M_K^2-2m_{\ell}^2+s_{\pi})\Delta_{\pi K}^2-\dfrac{2}{t_{\pi}^3}\,\Delta_{\ell K}\Delta_{\pi K}^3\,\bigg ]
\nonumber \\ 
&+& \dfrac{1}{16F_0^2}\,\dfrac{1}{s_{\pi}}\,B(t_{\pi},M_{\eta}^2,M_K^2)\,\bigg [\,\dfrac{2}{t_{\pi}}\,M_{\eta}^2s_{\pi}-\dfrac{2}{t_{\pi}^2}\,\Delta_{\eta K}^2s_{\pi}
\nonumber \\ 
&& +\dfrac{4}{t_{\pi}^2}\,(3M_{\eta}^2-2M_K^2)\Delta_{\ell K}\Delta_{\eta K}+\dfrac{4}{t_{\pi}^3}\,\Delta_{\ell K}\Delta_{\pi K}\Delta_{\eta K}^2\,\bigg ]\,,
\\ 
U_{\pi}^g
&=& \dfrac{8}{F_0^2}\,L_4
\nonumber \\ 
&+& \dfrac{1}{768\pi^2F_0^2}\,\dfrac{1}{s_{\pi}}\,\bigg\{\,
-12(M_{\eta}^2-4M_K^2+M_{\pi}^2-3s_{\pi})
\nonumber \\ 
&& -\dfrac{1}{t_{\pi}}\,\bigg [\,6(2M_K^2-3M_{\pi}^2-m_{\ell}^2)\Delta_{\pi K}+(M_K^2+5M_{\pi}^2)s_{\pi}
\nonumber \\ 
&& -18M_K^2(3M_{\eta}^2-8M_K^2-m_{\ell}^2)+3(M_{\eta}^2+M_K^2)s_{\pi}
\nonumber \\ 
&& \qquad -6M_{\eta}^2(7M_{\eta}^2+2M_K^2+m_{\ell}^2)\,\bigg ]
\nonumber \\ 
&& +\dfrac{1}{t_{\pi}^2}\,\bigg [\,4\Delta_{\pi K}^2s_{\pi}
\nonumber \\ 
&& +3M_K^2(7M_K^2-M_{\pi}^2+3m_{\ell}^2)\Delta_{\pi K}
\nonumber \\ 
&& -3M_{\pi}^2(2M_K^2-2M_{\pi}^2+m_{\ell}^2)\Delta_{\pi K}
\nonumber \\ 
&& -9M_K^2(10M_{\eta}^2-5M_K^2-m_{\ell}^2)\Delta_{\eta K}
\nonumber \\ 
&& +3M_{\eta}^2(66M_{\eta}^2-81M_K^2-13m_{\ell}^2)\Delta_{\eta K}\,\bigg ]
\nonumber \\ 
&& -\dfrac{6}{t_{\pi}^3}\,\bigg [\,(M_K^2+2M_{\pi}^2+m_{\ell}^2)\Delta_{\pi K}^3
\nonumber \\ 
&& +3(6M_{\eta}^2-9M_K^2-m_{\ell}^2)\Delta_{\eta K}^3\,\bigg ]\,\bigg\}
\nonumber \\ 
&+& \dfrac{1}{24F_0^2}\,\dfrac{1}{s_{\pi}}\,A(M_{\pi}^2)\,\bigg [\,-\dfrac{3}{M_{\pi}^2}\,s_{\pi}-\dfrac{2}{t_{\pi}}\,(3M_{\pi}^2+2s_{\pi})
\nonumber \\ 
&& -\dfrac{6}{t_{\pi}^2}\,(2M_K^2-3M_{\pi}^2)M_K^2
+\dfrac{3}{t_{\pi}^2}\,(3M_K^2-2M_{\pi}^2+m_{\ell}^2)M_{\pi}^2
\nonumber \\ 
&& -\dfrac{1}{t_{\pi}^2}\left( 7-\dfrac{2M_K^2}{M_{\pi}^2}\right) \Delta_{\pi K}s_{\pi}+\dfrac{6}{t_{\pi}^3}\,(M_K^2+2M_{\pi}^2+m_{\ell}^2)\Delta_{\pi K}^2\,\bigg ]
\nonumber \\ 
&+& \dfrac{1}{16F_0^2}\,\dfrac{1}{s_{\pi}}\,A(M_{\eta}^2)\times
\nonumber \\ 
&& \bigg [\,-\dfrac{4}{t_{\pi}}\,(5M_{\eta}^2-6M_K^2-s_{\pi})
-\dfrac{2}{t_{\pi}^2}\,\Delta_{\eta K}s_{\pi}
\nonumber \\ 
&& \qquad -\dfrac{2}{t_{\pi}^2}\,(18M_{\eta}^2-17M_K^2-5m_{\ell}^2)M_{\eta}^2
\nonumber \\ 
&& \qquad +\dfrac{4}{t_{\pi}^2}\,(20M_{\eta}^2-21M_K^2-3m_{\ell}^2)M_K^2
\nonumber \\ 
&& \qquad -\dfrac{12}{t_{\pi}^3}\,(M_K^2+2M_{\pi}^2+m_{\ell}^2)\Delta_{\eta K}^2\,\bigg ]
\nonumber \\ 
&+& \dfrac{1}{48F_0^2}\,\dfrac{1}{s_{\pi}}\,A(M_K^2)\times
\nonumber \\ 
&& \bigg [\,-\dfrac{2}{t_{\pi}}\,(6M_K^2-12M_{\pi}^2-7s_{\pi})
+\dfrac{6}{t_{\pi}}\,(4M_{\eta}^2-6M_K^2-s_{\pi})
\nonumber \\ 
&& +\dfrac{2}{t_{\pi}^2}\,(5\Delta_{\pi K}s_{\pi}-6\Sigma_{\ell K}M_{\pi}^2-15\Delta_{\pi K}M_K^2+3\Delta_{\ell \pi}M_K^2)
\nonumber \\ 
&& +\dfrac{6}{t_{\pi}^2}\,\Delta_{\eta K}s_{\pi}-\dfrac{6}{t_{\pi}^2}\,(3M_K^2+2m_{\ell}^2)M_{\eta}^2
\nonumber \\ 
&& \qquad +\dfrac{18}{t_{\pi}^2}\,(5M_K^2+m_{\ell}^2-3M_{\eta}^2)M_K^2
\nonumber \\ 
&& \qquad -\dfrac{12}{t_{\pi}^3}\,(M_K^2+2M_{\pi}^2+m_{\ell}^2)\Delta_{\pi K}^2
\nonumber \\ 
&& \qquad -\dfrac{36}{t_{\pi}^3}\,(6M_{\eta}^2-9M_K^2-m_{\ell}^2)\Delta_{\eta K}^2\,\bigg ]
\nonumber \\ 
&+& \dfrac{1}{768\pi^2F_0^2}\,\dfrac{1}{s_{\pi}}\,\ln\left( \dfrac{M_{\pi}^2}{M_K^2}\right) \,\bigg\{\,-2(3M_K^2-3M_{\pi}^2+2s_{\pi})
\nonumber \\ 
&& +\dfrac{1}{t_{\pi}}\left[ 2(2M_K^2+3M_{\pi}^2)s_{\pi}-3(3M_K^2+m_{\ell}^2)\Delta_{\pi K}\right] 
\nonumber \\ 
&& +\dfrac{1}{t_{\pi}^2}\left[ 2M_K^2s_{\pi}+3(M_K^2-4M_{\pi}^2-m_{\ell}^2)\Delta_{\pi K}\right] \Delta_{\pi K}
\nonumber \\ 
&& \qquad +\dfrac{1}{t_{\pi}^3}\,(9M_K^2+3m_{\ell}^2-2s_{\pi})\Delta_{\pi K}^3
\nonumber \\ 
&& \qquad +\dfrac{3}{t_{\pi}^4}\,(M_K^2+2M_{\pi}^2+m_{\ell}^2)\Delta_{\pi K}^4\,\bigg\}
\nonumber \\ 
&-& \dfrac{1}{512\pi^2F_0^2}\,\ln\left( \dfrac{M_{\eta}^2}{M_K^2}\right) \Delta_{\eta K}\times
\nonumber \\ 
&& \left[ \dfrac{2}{t_{\pi}}+\dfrac{2}{t_{\pi}^2}\,(M_{\eta}^2-3M_K^2)
-\dfrac{2}{t_{\pi}^3}\,(5M_{\eta}^2-3M_K^2)\Delta_{\eta K}
+\dfrac{6}{t_{\pi}^4}\,\Delta_{\eta K}^3\right] 
\nonumber \\ 
&-& \dfrac{1}{192\pi^2F_0^2}\,(4M_{\pi}^2-s_{\pi})\tau (s_{\pi},M_{\pi}^2,M_{\pi}^2)
\nonumber \\ 
&+& \dfrac{1}{768\pi^2F_0^2}\,\dfrac{1}{s_{\pi}}\,\tau (t_{\pi},M_{\pi}^2,M_K^2)(-\Delta_{\pi K}+t_{\pi})\times
\nonumber \\ 
&& \bigg\{\,2(3\Sigma_{\pi K}-2s_{\pi})+\dfrac{3}{t_{\pi}^4}\,(M_K^2+2M_{\pi}^2+m_{\ell}^2)\Delta_{\pi K}^4
\nonumber \\ 
&& -\dfrac{1}{t_{\pi}^3}\left[ 2\Delta_{\pi K}s_{\pi}-3M_{\pi}^2\Delta_{\ell K}+3M_K^2(5M_K^2+3m_{\ell}^2)\right] \Delta_{\pi K}^2
\nonumber \\ 
&& -\dfrac{1}{t_{\pi}^2}\left[ 3M_{\pi}^2(\Delta_{\ell K}+4M_{\pi}^2)+M_K^2(27M_K^2+9m_{\ell}^2-2s_{\pi})\right] \Delta_{\pi K}
\nonumber \\ 
&& -\dfrac{1}{t_{\pi}}\left[ 3M_{\pi}^2(\Delta_{\ell K}-2s_{\pi})+M_K^2(21M_K^2+3m_{\ell}^2-4s_{\pi})\right] \,\bigg\}
\nonumber \\ 
&+& \dfrac{1}{512\pi^2F_0^2}\,\tau (t_{\pi},M_{\eta}^2,M_K^2)(\Delta_{\eta K}^2-\Sigma_{\eta K}t_{\pi})\times
\nonumber \\ 
&& \left[ \dfrac{2}{t_{\pi}}+\dfrac{2}{t_{\pi}^2}\,(M_{\eta}^2-3M_K^2)
-\dfrac{2}{t_{\pi}^3}\,(5M_{\eta}^2-3M_K^2)\Delta_{\eta K}+\dfrac{6}{t_{\pi}^4}\,\Delta_{\eta K}^3\right] 
\nonumber \\ 
&+& \dfrac{1}{3F_0^2}\,B(s_{\pi},M_{\pi}^2,M_{\pi}^2)
\nonumber \\ 
&+& \dfrac{1}{24F_0^2}\,\dfrac{1}{s_{\pi}}\,B(t_{\pi},M_{\pi}^2,M_K^2)\,
\bigg\{\,-2s_{\pi}
\nonumber \\ 
&& +\dfrac{1}{t_{\pi}}\left[ 6\Delta_{\pi K}^2-(7M_K^2-4M_{\pi}^2)s_{\pi}\right] 
\nonumber \\ 
&& -\dfrac{1}{t_{\pi}^2}\left( 15M_K^2-6M_{\pi}^2+3m_{\ell}^2-7s_{\pi}\right) \Delta_{\pi K}^2
\nonumber \\ 
&& \quad -\dfrac{6}{t_{\pi}^3}\,(M_K^2+2M_{\pi}^2+m_{\ell}^2)\Delta_{\pi K}^3\,\bigg\}
\nonumber \\ 
&+& \dfrac{1}{16F_0^2}\,\dfrac{1}{s_{\pi}}\,B(t_{\pi},M_{\eta}^2,M_K^2)\times
\nonumber \\ 
&& \bigg\{\,\dfrac{2}{t_{\pi}}\left[ 2(5M_{\eta}^2-3M_K^2)\Delta_{\eta K}-(2M_{\eta}^2-M_K^2)s_{\pi}\right] 
\nonumber \\ 
&& -\dfrac{2}{t_{\pi}^2}\,\bigg [\,-M_{\eta}^2(18M_{\eta}^2-9M_K^2-5m_{\ell}^2)
\nonumber \\ 
&& -\Delta_{\eta K}s_{\pi}+3M_K^2(10M_{\eta}^2-5M_K^2-m_{\ell}^2)\,\bigg ]\Delta_{\eta K}
\nonumber \\
&& \quad +\dfrac{12}{t_{\pi}^3}\,(M_K^2+2M_{\pi}^2+m_{\ell}^2)\Delta_{\eta K}^3\,\bigg\}\,.
\end{eqnarray} 
In the preceding expressions, 
\begin{equation}
\Delta_{\ell P}\,\doteq\,m_{\ell}^2-M_P^2\,, \qquad \Sigma_{\ell P}\,\doteq\,m_{\ell}^2+M_P^2\,,
\end{equation} 
\begin{equation}
\Delta_{PQ}\,\doteq\,M_P^2-M_Q^2\,, \qquad \Sigma_{PQ}\,\doteq\,M_P^2+M_Q^2\,.
\end{equation} 

\subsection{The $\Delta_K$-terms}

\begin{eqnarray}
U_K^f
&=& \dfrac{1}{768\pi^2F_0^2}\,\bigg\{\,-16\,\dfrac{M_{\pi}^2+2M_K^2}{M_{\pi}^2-M_{\eta}^2}
\nonumber \\ 
&& \quad +12+\dfrac{3}{t_{\pi}}\,(3M_{\eta}^2-6M_K^2-M_{\pi}^2)
\nonumber \\ 
&& +\dfrac{2}{t_{\pi}^2}\left[ 2(M_K^2-3M_{\pi}^2)\Delta_{\pi K}-9\Delta_{\eta K}^2\right] -\dfrac{4}{t_{\pi}^3}\,\Delta_{\pi K}^3\,\bigg\}
\nonumber \\ 
&+& \dfrac{1}{24F_0^2}\,\dfrac{1}{t_{\pi}^2}\,\Delta_{\pi K}A(M_{\pi}^2)
\nonumber \\ 
&-& \dfrac{1}{24F_0^2}\,\dfrac{1}{M_K^2}\,A(M_K^2)\times
\nonumber \\ 
&& \bigg\{\,-8\,\dfrac{M_{\pi}^2+2M_K^2}{M_{\pi}^2-M_{\eta}^2}
+\dfrac{1}{t_{\pi}}\,(9M_{\eta}^2-13M_K^2-5M_{\pi}^2)
\nonumber \\ 
&& +\dfrac{1}{t_{\pi}^2}\left[ (2M_K^2-M_{\pi}^2)\Delta_{\pi K}+9(2M_K^2-M_{\eta}^2)\Delta_{\eta K}\right] \,\bigg\}
\nonumber \\ 
&+& \dfrac{1}{8F_0^2}\left( -\dfrac{1}{t_{\pi}}+\dfrac{3}{t_{\pi}^2}\,\Delta_{\eta K}\right) A(M_{\eta}^2)
\nonumber \\ 
&+& \dfrac{1}{384\pi^2F_0^2}\,\ln\left( \dfrac{M_{\pi}^2}{M_K^2}\right) \,\bigg [\,-3+\dfrac{1}{t_{\pi}}\,(4M_K^2-M_{\pi}^2)
\nonumber \\ 
&& +\dfrac{1}{t_{\pi}^2}\,M_K^2\Delta_{\pi K}-\dfrac{1}{t_{\pi}^3}\,(M_K^2-3M_{\pi}^2)\Delta_{\pi K}^2+\dfrac{1}{t_{\pi}^4}\,\Delta_{\pi K}^4\,\bigg ]
\nonumber \\ 
&+& \dfrac{1}{512\pi^2F_0^2}\,\ln\left( \dfrac{M_{\eta}^2}{M_K^2}\right) \,\bigg [\,\dfrac{2}{t_{\pi}}\,(3M_{\eta}^2-M_K^2)
\nonumber \\ 
&& -\dfrac{4}{t_{\pi}^2}\,(3M_{\eta}^2-2M_K^2)\Delta_{\eta K}+\dfrac{6}{t_{\pi}^3}\,\Delta_{\eta K}^3\,\bigg ]
\nonumber \\ 
&+& \dfrac{1}{384\pi^2F_0^2}\,\tau (t_{\pi},M_{\pi}^2,M_K^2)\,\bigg [\,-(7M_K^2-M_{\pi}^2-3t_{\pi})
\nonumber \\ 
&& -\dfrac{1}{t_{\pi}}\,(5M_K^2+3M_{\pi}^2)\Delta_{\pi K}+\dfrac{2}{t_{\pi}^2}\,M_{\pi}^2\Sigma_{\pi K}\Delta_{\pi K}-\dfrac{2}{t_{\pi}^3}\,\Delta_{\pi K}^4-\dfrac{1}{t_{\pi}^4}\,\Delta_{\pi K}^5\,\bigg ]
\nonumber \\ 
&-& \dfrac{1}{512\pi^2F_0^2}\,\tau (t_{\pi},M_{\eta}^2,M_K^2)(\Delta_{\eta K}+t_{\pi})\times
\nonumber \\ 
&& \bigg [\,\dfrac{2}{t_{\pi}}\,(3M_{\eta}^2-M_K^2)
-\dfrac{4}{t_{\pi}^2}\,(3M_{\eta}^2-2M_K^2)\Delta_{\eta K}+\dfrac{6}{t_{\pi}^3}\,\Delta_{\eta K}^3\,\bigg ]
\nonumber \\ 
&+& \dfrac{1}{24F_0^2}\,B(t_{\pi},M_{\pi}^2,M_K^2)\left[ 3+\dfrac{1}{t_{\pi}}\,(2M_K^2-5M_{\pi}^2)-\dfrac{2}{t_{\pi}^2}\,\Delta_{\pi K}^2\right] 
\nonumber \\ 
&+& \dfrac{1}{16F_0^2}\,B(t_{\pi},M_{\eta}^2,M_K^2)\left[ 4+\dfrac{2}{t_{\pi}}\,(4M_{\eta}^2-5M_K^2)-\dfrac{12}{t_{\pi}^2}\,\Delta_{\eta K}^2\right]\,,
\\ 
U_K^g
&=& \dfrac{2}{F_0^2}\,(L_3+2L_4)
\nonumber \\ 
&+& \dfrac{1}{768\pi^2F_0^2}\,\bigg\{\,6
+\dfrac{1}{t_{\pi}}\,(34M_K^2+9M_{\eta}^2-3M_{\pi}^2)
\nonumber \\ 
&& +\dfrac{1}{t_{\pi}^2}\left[ (17M_K^2-3M_{\pi}^2)\Delta_{\pi K}-3(7M_{\eta}^2+3M_K^2)\Delta_{\eta K}\right] 
\nonumber \\ 
&& \qquad -\dfrac{2}{t_{\pi}^3}\,(\Delta_{\pi K}^3-9\Delta_{\eta K}^3)\,\bigg\}
\nonumber \\ 
&+& \dfrac{1}{8F_0^2}\,A(M_{\pi}^2)\left[ \dfrac{1}{t_{\pi}}-\dfrac{1}{t_{\pi}^2}\,(3M_K^2-4M_{\pi}^2)
+\dfrac{2}{t_{\pi}^3}\,\Delta_{\pi K}^2\right] 
\nonumber \\ 
&+& \dfrac{1}{16F_0^2}\,A(M_{\eta}^2)\left[ \dfrac{2}{t_{\pi}^2}\,(4M_{\eta}^2-5M_K^2)-\dfrac{12}{t_{\pi}^3}\,\Delta_{\eta K}^2\right] 
\nonumber \\ 
&-& \dfrac{1}{48F_0^2}\,A(M_K^2)\,\bigg\{\,-\dfrac{4}{M_K^2}+\dfrac{3}{t_{\pi}}\left( 22-\dfrac{4M_{\pi}^2}{M_K^2}-\dfrac{4M_{\eta}^2}{M_K^2}\right) 
\nonumber \\ 
&& -\dfrac{3}{t_{\pi}^2}\left[ 8M_K^2-2M_{\pi}^2\left( 7-\dfrac{M_{\pi}^2}{M_K^2}\right) +2M_{\eta}^2\left( 5-\dfrac{3M_{\eta}^2}{M_K^2}\right) \right] 
\nonumber \\ 
&& \qquad +\dfrac{12}{t_{\pi}^3}\,(\Delta_{\pi K}^2-3\Delta_{\eta K}^2)\,\bigg\}
\nonumber \\ 
&-& \dfrac{1}{768\pi^2F_0^2}\,\ln\left( \dfrac{M_{\pi}^2}{M_K^2}\right) (\Delta_{\pi K}+3t_{\pi})\times
\nonumber \\ 
&& \left[ \dfrac{1}{t_{\pi}}+\dfrac{2}{t_{\pi}^2}\,M_K^2
+\dfrac{4}{t_{\pi}^3}\,M_K^2\Delta_{\pi K}-\dfrac{1}{t_{\pi}^4}\,\Delta_{\pi K}^3\right] 
\nonumber \\ 
&-& \dfrac{1}{512\pi^2F_0^2}\,\ln\left( \dfrac{M_{\eta}^2}{M_K^2}\right) (\Delta_{\eta K}+t_{\pi})\times
\nonumber \\ 
&& \left[ \dfrac{2}{t_{\pi}}+\dfrac{2}{t_{\pi}^2}\,(M_{\eta}^2-3M_K^2)
-\dfrac{2}{t_{\pi}^3}\,(5M_{\eta}^2-3M_K^2)\Delta_{\eta K}+\dfrac{6}{t_{\pi}^4}\,\Delta_{\eta K}^3\right] 
\nonumber \\ 
&-& \dfrac{1}{384\pi^2F_0^2}\,(4M_K^2-s_{\pi})\tau (s_{\pi},M_K^2,M_K^2)
\nonumber \\ 
&+& \dfrac{1}{768\pi^2F_0^2}\,\tau (t_{\pi},M_{\pi}^2,M_K^2)\,\bigg [\,2M_K^2+8M_{\pi}^2+3t_{\pi}
\nonumber \\ 
&& \qquad -\dfrac{1}{t_{\pi}}\,(19M_K^4-8M_{\pi}^2M_K^2+5M_{\pi}^4)
\nonumber \\ 
&& \quad -\dfrac{1}{t_{\pi}^2}\,(21M_K^4-16M_{\pi}^2M_K^2+3M_{\pi}^4)\Delta_{\pi K}
\nonumber \\ 
&& \quad +\dfrac{2}{t_{\pi}^3}\,(4M_K^2-M_{\pi}^2)\Delta_{\pi K}^3-\dfrac{1}{t_{\pi}^4}\,\Delta_{\pi K}^5\,\bigg ]
\nonumber \\ 
&+& \dfrac{1}{512\pi^2F_0^2}\,\tau (t_{\pi},M_{\eta}^2,M_K^2)\left( -2M_K^2+t_{\pi}+\dfrac{1}{t_{\pi}}\,\Delta_{\eta K}^2\right) \times
\nonumber \\ 
&& \left[ 2+\dfrac{2}{t_{\pi}}\,(M_{\eta}^2-3M_K^2)
-\dfrac{2}{t_{\pi}^2}\,(5M_{\eta}^2-3M_K^2)\Delta_{\eta K}+\dfrac{6}{t_{\pi}^3}\,\Delta_{\eta K}^3\right] 
\nonumber \\ 
&+& \dfrac{1}{6F_0^2}\,B(s_{\pi},M_K^2,M_K^2)
\nonumber \\ 
&+& \dfrac{1}{8F_0^2}\,B(t_{\pi},M_{\pi}^2,M_K^2)\times
\nonumber \\ 
&& \left[ -2+\dfrac{3}{t_{\pi}}\,(2M_K^2-M_{\pi}^2)-\dfrac{5}{t_{\pi}^2}\,\Delta_{\pi K}^2-\dfrac{2}{t_{\pi}^3}\,\Delta_{\pi K}^3\right] 
\nonumber \\ 
&+& \dfrac{1}{16F_0^2}\,B(t_{\pi},M_{\eta}^2,M_K^2)\times
\nonumber \\ 
&& \left[ 2-\dfrac{2}{t_{\pi}}\,(2M_{\eta}^2-5M_K^2)-\dfrac{2}{t_{\pi}^2}\,\Sigma_{\eta K}\Delta_{\eta K}+\dfrac{12}{t_{\pi}^3}\,\Delta_{\eta K}^3\right] \,.
\end{eqnarray}

%% file: section_6.tex
\section{Results}
\label{sec:results}

\subsection{Input}
\label{sec:input}

The numerical values of the physical parameters must be fixed through experimental input. However, this input may not necessarily consist of direct measurements of the renormalized parameters; it may be obtained from any suitable set of experimental results. In practice one uses those experiments which have the highest experimental accuracy and theoretical reliability. This criterion is certainly fulfilled for the following set of parameters whose numerical values are taken from~\cite{Hagiwara:2002fs}:
\begin{itemize}
\item the fine structure constant,
$$
\alpha\,=\,1/137.03599976(50)\,,
$$
corresponding to the classical electron charge $e\,=\,\sqrt{4\pi\alpha}$,
\item the masses of the charged leptons,
$$
m_{\mathrm{e}}\,=\,0.510998902(21)~\mathrm{MeV}\,, \qquad m_{\mu}\,=\,105.658357(5)~\mathrm{MeV}\,,
$$
\item the Fermi constant,
$$
G_F\,=\,1.16639(1)\cdot 10^{-5}~\mathrm{GeV}^{-2}\,,
$$
which is directly related to the muon lifetime,
\item the Cabibbo-Kobayashi-Maskawa quark-mixing matrix element,
$$
|V_{us}|\,=\,0.2196\pm 0.0026\,,
$$
coming from the analysis of $K_{\mathrm{e}3}$ decays,
\item the masses of the light mesons,
$$
M_{\pi^{\pm}}\,=\,139.57018(35)~\mathrm{MeV}\,, \qquad M_{\pi^0}\,=\,134.9766(6)~\mathrm{MeV}\,,
$$
$$
M_{K^{\pm}}\,=\,493.677\pm 0.016~\mathrm{MeV}\,, \qquad M_{K^0}\,=\,497.672\pm 0.031~\mathrm{MeV}\,,
$$
$$
M_{\eta}\,=\,547.30\pm 0.12~\mathrm{MeV}\,, \qquad M_{\rho}\,=\,771.1\pm 0.9~\mathrm{MeV}\,,
$$
\item the charged light mesons decay constants,
$$
F_{\pi^{\pm}}\,=\,92.419\pm 0.325~\mathrm{MeV}\,, \qquad F_{K^{\pm}}\,=\,112.996\pm 1.301~\mathrm{MeV}\,,
$$
coming from the analysis of $\pi_{\mu 2}$ and $K_{\mu 2}$ decays, respectively.
\end{itemize}

Let us consider the parameters, $M_{\pi}$, $M_K$ and $\epsilon$,
related to light quark masses. Since $M_{\pi}$ and $M_K$ figure in
our expressions only at next-to-leading order, it is completely
safe to replace them by their leading order expressions. In fact,
the quantity $M_{\pi}$ will be identified to the neutral pion
mass,
$$
M_{\pi}\,\longrightarrow\,M_{\pi^0}\,=\,134.9766(6)~\mathrm{MeV}\,,
$$
while $M_K^2$ will be replaced by,
$$
M_K^2\,\longrightarrow\,\frac{1}{2}\,\left( M_{K^{\pm}}^2+M_{K^0}^2+M_{\pi^0}^2-M_{\pi^{\pm}}^2\right) \,,
$$
to get,
$$
M_K\,=\,495.042\pm 0.034~\mathrm{MeV}\,.
$$
For $\epsilon$, we will use the value~\cite{Leutwyler:1996qg},
$$
\epsilon\,=\,(1.061\pm 0.083)\cdot 10^{-2}\,.
$$
extracted from the mass splitting in the baryon octet.

We will turn now to the determination of low-energy constants in the strong sector. Following~\cite{Gasser:1985gg}, these constants will be evaluated at one-loop accuracy, that is, by fitting experimental measurements of the concerned observables to their ChPT expressions at next-to-leading order. Note that all of our expressions will be evaluated at the scale $\mu$ equal to the rho mass. The $K_{\ell 4}$ form factors are sensitive to variations of the low-energy constants, $L_1$, $L_2$ and $L_3$. By fitting experimental results on $K_{\ell 4}$ form factors~\cite{Rosselet:1977pu} to their ChPT expressions at next-to-leading chiral order we obtain~\cite{Amoros:1999qq},
$$
L_1^r\,=\,(0.46\pm 0.24)\cdot 10^{-3}\,,
$$
$$
L_2^r\,=\,(1.49\pm 0.23)\cdot 10^{-3}\,, \; L_3^r\,=\,(-3.18\pm 0.85)\cdot 10^{-3}\,.
$$
The constant $L_5$ can be extracted from the ratio of the kaon to the pion decay constant in the isospin limit~\cite{Gasser:1985gg},
\begin{eqnarray}
\dfrac{F_{K^{\pm}}}{F_{\pi^{\pm}}}
&=& 1+\dfrac{4}{F_{\pi^{\pm}}^2}\,(M_{K^{\pm}}^2-M_{\pi^{\pm}}^2)L_5^r
+\dfrac{5M_{\pi^{\pm}}^2}{128\pi^2F_{\pi^{\pm}}^2}\,\ln\dfrac{M_{\pi^{\pm}}^2}{\mu^2}
\nonumber \\
&-& \dfrac{M_{K^{\pm}}^2}{64\pi^2F_{\pi^{\pm}}^2}\,\ln\dfrac{M_{K^{\pm}}^2}{\mu^2}
-\dfrac{3M_{\eta}^2}{128\pi^2F_{\pi^{\pm}}^2}\,\ln\dfrac{M_{\eta}^2}{\mu^2}\,,
\nonumber
\end{eqnarray}
and reads,
$$
L_5^r\,=\,(1.49\pm 0.14)\cdot 10^{-3}\,.
$$
Having $L_5$, it is easy to determine $L_8$ from the quantity $\Delta_M$ accounting for $SU(3)$ breaking~\cite{Gasser:1985gg},
\begin{eqnarray}
\Delta_M
&=& \dfrac{8}{F_{\pi^{\pm}}^2}\,(M_{K^{\pm}}^2-M_{\pi^{\pm}}^2)(2L_8^r-L_5^r)
\nonumber \\
&-& \dfrac{M_{\pi^{\pm}}^2}{32\pi^2F_{\pi^{\pm}}^2}
\,\ln\dfrac{M_{\pi^{\pm}}^2}{\mu^2}
+\dfrac{M_{\eta}^2}{32\pi^2F_{\pi^{\pm}}^2}\,\ln\dfrac{M_{\eta}^2}{\mu^2}\,,
\nonumber
\end{eqnarray}
and which value reads~\cite{Leutwyler:1996qg},
$$
\Delta_M\,=\,0.065\pm 0.065\,.
$$
The result is,
$$
L_8^r\,=\,(1.02\pm 0.22)\cdot 10^{-3}\,.
$$
The constant $L_7$ is obtained from $L_5$ and $L_8$ with the help of the isospin limit quantity,
\begin{equation}
\Delta_{\mathrm{GMO}}\,\doteq\,(4M_{K^{\pm}}^2-M_{\pi^{\pm}}^2-3M_{\eta}^2)
/(M_{\eta}^2-M_{\pi^{\pm}}^2)\,=\,0.2027(15)\,,
\end{equation}
by matching its value to the ChPT expression at next-to-leading order~\cite{Gasser:1985gg},
\begin{eqnarray}
\Delta_{\mathrm{GMO}}
&=& -\dfrac{6}{F_{\pi^{\pm}}^2}\,(M_{\eta}^2-M_{\pi^{\pm}}^2)(12L_7+6L_8^r-L_5^r)
\nonumber \\
&+& \dfrac{2}{M_{\eta}^2-M_{\pi^{\pm}}^2}\left( \dfrac{M_{\pi^{\pm}}^4}{32\pi^2F_{\pi^{\pm}}^2}
\,\ln\dfrac{M_{\pi^{\pm}}^2}{\mu^2}\right.
\nonumber \\
&-& \left.\dfrac{M_{K^{\pm}}^4}{8\pi^2F_{\pi^{\pm}}^2}
\,\ln\dfrac{M_{K^{\pm}}^2}{\mu^2}+\dfrac{3M_{\eta}^4}{32\pi^2F_{\pi^{\pm}}^2}
\,\ln\dfrac{M_{\eta}^2}{\mu^2}\right)\,.
\end{eqnarray}
We obtain for $L_7$ the value,
$$
L_7\,=\,(-0.44\pm 0.12)\cdot 10^{-3}\,.
$$
The constant $L_9$ is fixed from the electromagnetic charge radius of the pion~\cite{Bijnens:2002hp},
$$
L_9^r\,=\,(5.5\pm 0.2)\cdot 10^{-3}\,.
$$
Finally, it is difficult to fix the constants $L_4$ and $L_6$ by direct experimental determination. These constants are suppressed by the Okubo-Zweig-Iizuka (OZI) rule and measure the amount by which $m_s$ affects the values of the order parameters $F$ and $\langle\overline{q}q\rangle$. The constant $L_4$ was derived from Roy and Steiner equations for $S$- and $P$-waves of $\pi -K$ scattering amplitude~\cite{Buettiker:2003pp},
$$
L_4^r\,=\,(0.53\pm 0.39)\cdot 10^{-3}\,.
$$
The constant $L_6$ has been obtained from a chiral sum rule~\cite{Moussallam:1999aq},
$$
L_6^r\,=\,(0.4\pm 0.2)\cdot 10^{-3}\,.
$$

To close the discussion about the strong sector we have to fix the
parameter $F_0$. At leading chiral order this parameter is given
by the pseudoscalar decay constants, $F_{\pi}$, $F_K$ or
$F_{\eta}$. One can then see the latter as the ``renormalized''
quantities corresponding to the ``bare'' quantity $F_0$ and thus
replace it by one of them after accounting for next-to-leading
order contributions. But the main question is which expression for
the decay constants to use especially that the difference between
their numerical values is relatively big. For instance, the
expressions for the pion and kaon decay constants at
next-to-leading order are given in the isospin limit
by~\cite{Gasser:1985gg},
\begin{eqnarray}
F_{\pi^{\pm}}
&=& F_0\left[ 1+\dfrac{4}{F_{\pi^{\pm}}^2}\,(M_{\pi^{\pm}}^2+2M_{K^{\pm}}^2)L_4^r
+\dfrac{4M_{\pi^{\pm}}^2}{F_{\pi^{\pm}}^2}\,L_5^r\right.
\nonumber \\
&-& \left.\dfrac{M_{\pi^{\pm}}^2}{16\pi^2F_{\pi^{\pm}}^2}\,
\ln\dfrac{M_{\pi^{\pm}}^2}{\mu^2}-\dfrac{M_{K^{\pm}}^2}{32\pi^2F_{\pi^{\pm}}^2}\,
\ln\dfrac{M_{K^{\pm}}^2}{\mu^2}\right] \,,
\nonumber \\
F_{K^{\pm}}
&=& F_0\left[ 1+\dfrac{4}{F_{\pi^{\pm}}^2}\,(M_{\pi^{\pm}}^2+2M_{K^{\pm}}^2)L_4^r
+\dfrac{4M_{K^{\pm}}^2}{F_{\pi^{\pm}}^2}\,L_5^r\right.
\nonumber \\
&-& \left.\dfrac{3M_{\pi^{\pm}}^2}{128\pi^2F_{\pi^{\pm}}^2}\,
\ln\dfrac{M_{\pi^{\pm}}^2}{\mu^2}-\dfrac{3M_{K^{\pm}}^2}{64\pi^2F_{\pi^{\pm}}^2}\,
\ln\dfrac{M_{K^{\pm}}^2}{\mu^2}-\dfrac{3M_{\eta}^2}{128\pi^2F_{\pi^{\pm}}^2}\,
\ln\dfrac{M_{\eta}^2}{\mu^2}\right] \,.
\nonumber
\end{eqnarray}
Taking as input the aforecited values for $M_{\pi^{\pm}}$, $M_{K^{\pm}}$, $F_{\pi^{\pm}}$, $F_{K^{\pm}}$, $L_4^r$ and $L_5^r$, we obtain for $F_0$ the central values, $F_0=67.53$ MeV and $F_0=57.40$ MeV from $F_{\pi^{\pm}}$ and $F_{K^{\pm}}$, respectively. If, for comparison, we take for $L_4$ its large-$N_c$ estimate, the central values modify to $F_0=79.16$ MeV and $F_0=71.62$ MeV from $F_{\pi^{\pm}}$ and $F_{K^{\pm}}$, respectively. This amounts for a $15\%$ to $20\%$ deviation for the value of $F_0$. In our calculation we will use for $F_0$ the two values given by the bounds of the following inequality,
$$
57.40\,\leq\,F_0\,\leq\,67.53\,,
$$
and give the difference between the two obtained results as an error on the final result.

In the electroweak sector it is quasi impossible to have an
experimental determination of the low-energy constants due to the
relatively big number of constants from one side and to the
relatively small magnitude of the electroweak effects from the
other side. We will use for the constants $K_i$ in the mesonic
sector the following central values obtained by means of resonance
saturation~\cite{Baur:1997ya},
$$
\begin{array}{ccc}
K_1^r\,=-6.4\cdot 10^{-3}\,, & K_2^r\,=-3.1\cdot 10^{-3}\,, & K_3^r\,=6.4\cdot 10^{-3}\,, \\
K_4^r\,=-6.4\cdot 10^{-3}\,, & K_5^r\,=19.9\cdot 10^{-3}\,, & K_6^r\,=8.6\cdot 10^{-3}\,, \\
K_9^r\,=0\,,                 & K_{10}^r\,=0\,,              & K_{12}^r\,=-9.2\cdot 10^{-3}\,,
\end{array}
$$
with an error of $\pm 6.3\cdot 10^{-3}$ assigned to each of them coming from na\"ive dimensional analysis. The latter will also be used to fix the bounds on low-energy constants in the electroweak leptonic sector,
$$
|X_i|\,\leq\,6.3\cdot 10^{-3}\,,
$$
since these constants have not been yet determined.

\subsection{The $f$ form factor}

In what follows we will refer to the inequality,
\begin{equation}
4M_{\pi^{\pm}}^2<s_{\pi}<(M_{K^{\pm}}-m_{\ell})^2\,,
\end{equation}
as the \textit{allowed kinematical region}. The first term in the partial wave expansion for $f$ form factor is infrared finite. It contains however singular (Coulomb) terms for,
\begin{equation}
s_{\pi}\,=\,(M_K-m_{\ell})^2+2m_{\ell}(M_K\mp 2M_{\pi}-m_{\ell})\,.
\end{equation}
As can be easily seen, the singularity is outside the allowed kinematical region for $m_{\ell}\neq 0$ and approaches the upper bound from the right when $m_{\ell}$ tends to zero. Therefore, there is no apparent reason for subtracting Coulomb terms in the case of non-vanishing lepton mass. In order to see the impact of such terms on the whole correction, let us consider the following imaginary part,
\begin{eqnarray}
\mathrm{Im}\,U^f(s_{\pi})
&=& \dfrac{3}{32\pi F_0^2}\,\dfrac{\epsilon}{\sqrt{3}}\,(s_{\pi}-2M_{\pi}^2)\left( 1-\dfrac{4M_{\pi}^2}{s_{\pi}}\right) ^{1/2}
\nonumber \\
&& +\dfrac{\Delta_{\pi}}{48\pi F_0^2}\left( 2-\dfrac{5M_{\pi}^2}{s_{\pi}}\right) \left( 1-\dfrac{4M_{\pi}^2}{s_{\pi}}\right) ^{1/2}
\nonumber \\
&& +\dfrac{3e^2}{32\pi}\,\dfrac{m_{\ell}^2}{t_{\pi}}
\,(t_{\pi}+M_{\pi}^2-m_{\ell}^2)\lambda^{-1/2}(t_{\pi},m_{\ell}^2,M_{\pi}^2)\,.
\end{eqnarray}
The plot of the preceding expression as a function of $s_{\pi}$ is given by figure~\ref{fig:f-phase_electron}. It is easy to see that $e^2$ (singular) -terms are almost negligible with respect to $\Delta_{\pi}$ or $\epsilon$-terms.

\subsection{The $g$ form factor}

Unlike the $f$ form factor, the $g$ form factor is infrared divergent. We have shown in~\cite{Nehme:2003bz} that this divergence is cancelled at the level of differential decay rate by the one coming from real soft photon emission. In $K_{\ell 4}$ experiments, one has to measure modules and phases for form factors. Therefore, a subtraction of the infrared divergence should be applied at the level of form factors. The trouble is that the subtraction is not unique. A possible choice corresponds to a \textit{minimal subtraction} and consists on dropping out the $\ln m_{\gamma}$ term. Another possible choice which we qualify by a \textit{reasonable subtraction} consists on treating $f$ and $g$ form factors on a equal footing. While the $f$ form factor is infrared finite, the infrared divergence in $g$ form factor comes from wave function renormalization of external charged particles and from virtual photon exchange. The latter contribution is generated from the $C_0$ function,
\begin{equation}
C_0(-p_l,p_2,m_{\gamma},m_l,M_{\pi})\,,
\end{equation}
expressed by formula (195) in the appendix of
reference~\cite{Nehme:2003bz}. In the reasonable subtraction
scheme, one drops out the $\ln m_{\gamma}$ term coming from wave
function renormalization and the \textit{full} contribution of the
$C_0$ function. Formally, one introduces a subtraction parameter,
$\xi$, which equals $1$ in the minimal subtraction scheme and
vanishes in the reasonable one. Having this, we define the
subtracted real part,
\begin{eqnarray}
g_P(s_{\pi},\xi )
&=& 1+\mathrm{Re}\,U^g(s_{\pi})
\nonumber \\
&& +\dfrac{e^2}{8\pi^2}\,\ln m_{\gamma}^2
\nonumber \\
&& -\dfrac{e^2}{8\pi^2}\,\dfrac{t_{\pi}-M_{\pi}^2-m_{\ell}^2}
{\sqrt{t_{\pi}-(m_{\ell}+M_{\pi})^2}\sqrt{t_{\pi}-(m_{\ell}-M_{\pi})^2}}\,\xi\,\times
\nonumber \\
&& \ln\dfrac{\sqrt{t_{\pi}-(m_{\ell}-M_{\pi})^2}+\sqrt{t_{\pi}-(m_{\ell}+M_{\pi})^2}}
{\sqrt{t_{\pi}-(m_{\ell}-M_{\pi})^2}-\sqrt{t_{\pi}-(m_{\ell}+M_{\pi})^2}}\,
\ln m_{\gamma}^2
\nonumber \\
&& +2e^2(t_{\pi}-M_{\pi}^2-m_{\ell}^2)\times
\nonumber \\
&& (1-\xi )\,\mathrm{Re}\,C(m_{\ell}^2,t_{\pi},M_{\pi}^2,m_{\gamma}^2,m_{\ell}^2,M_{\pi}^2)\,.
\end{eqnarray}
Finally, from the imaginary part,
\begin{eqnarray}
\mathrm{Im}\,U^g(s_{\pi})
&=& \delta_1^1(s_{\pi})
\nonumber \\
&& +\dfrac{\Delta_{\pi}}{32\pi F_0^2}\left( 1-\dfrac{4M_{\pi}^2}{s_{\pi}}\right) ^{1/2}
\nonumber \\
&& +\dfrac{e^2}{32\pi t_{\pi}}\,\lambda^{-1/2}(t_{\pi},m_{\ell}^2,M_{\pi}^2)\times
\nonumber \\
&& \left[ m_{\ell}^2(5t_{\pi}+M_{\pi}^2-m_{\ell}^2)+4t_{\pi}(M_{\pi}^2-t_{\pi})\right]
\nonumber \\
&& -2e^2(t_{\pi}-M_{\pi}^2-m_{\ell}^2)\,
\mathrm{Im}\,C(m_{\ell}^2,t_{\pi},M_{\pi}^2,m_{\gamma}^2,m_{\ell}^2,M_{\pi}^2)\,,
\end{eqnarray}
where,
\begin{equation}
\delta_1^1(s_{\pi})\,=\,\dfrac{s_{\pi}}{96\pi F_0^2}\left( 1-\dfrac{4M_{\pi^{\pm}}^2}{s_{\pi}}\right) ^{3/2}\,,
\end{equation}
we define the subtracted phase as,
\begin{eqnarray}
\delta_P(s_{\pi},\xi )
&\doteq& \mathrm{Im}\,U^g(s_{\pi})
\nonumber \\
&& +\dfrac{e^2}{8\pi}\,(t_{\pi}-M_{\pi}^2-m_{\ell}^2)
\lambda^{-1/2}(t_{\pi},m_{\ell}^2,M_{\pi}^2)\,\xi\,\ln m_{\gamma}^2
\nonumber \\
&& +2e^2(t_{\pi}-M_{\pi}^2-m_{\ell}^2)\times
\nonumber \\
&& (1-\xi )\,\mathrm{Im}\,C(m_{\ell}^2,t_{\pi},M_{\pi}^2,m_{\gamma}^2,m_{\ell}^2,M_{\pi}^2)\,.
\end{eqnarray}

%% file: conclusion.tex
\section{Conclusion}

In this work we proposed a possible splitting between strong and electromagnetic interactions in $K_{\ell 4}$ decay form factors. The technique was applied to the decay of neutral kaon, $K^0\longrightarrow\pi^0\pi^-\ell^+\nu_{\ell}$. It consists on working at the production threshold for the lepton pair, $s_{\ell}=m_{\ell}^2$. The latter assumption simplifies significantly the splitting by allowing a partial wave expansion of form factors with exactly the same structure as in the pure strong theory. This constitutes a good approximation as long as the dependence of form factors on $s_{\ell}$ remains linear; the slope poor. 

The interest in the present process is at first theoretical. In fact, the partial wave expansion of form factors involves the $P$-wave iso-vector $\pi\pi$ phase shift, $\delta_1^1(s_{\pi})$, which can be related to $\pi\pi$ scattering lengths via Roy equations. In their turn, scattering lengths are sensitive to the way Chiral symmetry is spontaneously broken. Consequently, a theoretical study of the process in question including all possible contributions is imperative. We gave here the first analytic and numerical evaluation of the Isospin breaking contribution. This would allow the extraction of $\delta_1^1(s_{\pi})$ from the experimental measurement of form factors. 

We started with the evaluation of the first term in the partial wave expansion for $f$ form factor. This term vanishes in the absence of Isospin breaking and is free from Infrared divergences in its presence. Motivated by these two features, we studied the sensitivity of Isospin breaking correction to variations of $F_0$ and $m_{\ell}$. This was achieved by plotting the graph of the correction as a function of $s_{\pi}$ for two values of $F_0$ in figure~\ref{fig:f-module_electron} and for $m_{\ell}=m_{\mathrm{e}},m_{\mu}$ in figure~\ref{fig:f-module_mass}. We then compared in figure~\ref{fig:f-phase_electron} the relative size for the different contributions to the correction coming from virtual photons, $\mathcal{O}(e^2)$, mass square difference between charged and neutral mesons, $\mathcal{O}(Z_0e^2)$, and mass difference between up and down quarks, $\mathcal{O}(m_d-m_u)$. 

We pursued with the evaluation of the first term in the partial wave expansion for $g$ form factor. The comparison between the size of Isospin breaking correction to the real part of the term in question and the one-loop level correction to the same quantity and in the absence of Isospin breaking was made in figure~\ref{fig:g-module_all}. $\mathcal{O}(m_d-m_u)$ and $\mathcal{O}(\alpha )$ contributions to the preceding correction were compared in figure~\ref{fig:g-module_breaking}. Finally, Isospin breaking correction to the $P$-wave iso-vector $\pi\pi$ phase shift was plotted in figure~\ref{fig:g-phase_electron}. 

Our results are of great utility for the interpretation of the outgoing data from the KTeV experiment at FNAL. 

%% file: appendix.tex
\section{Loop Integrals}

\subsection{$B$-integrals}

\begin{equation}
B(M_{\pi}^2,0,M_{\pi}^2)\,=\,-2\overline{\lambda}
+\dfrac{1}{16\pi^2}\left[ 1-\ln\left( \dfrac{M_{\pi}^2}{\mu^2}\right) \right] \,.
\end{equation} 

\begin{equation}
B(m_{\ell}^2,0,m_{\ell}^2)\,=\,-2\overline{\lambda}
+\dfrac{1}{16\pi^2}\left[ 1-\ln\left( \dfrac{m_{\ell}^2}{\mu^2}\right) \right] \,.
\end{equation}

\begin{eqnarray}
B(0,m_{\ell}^2,M_K^2)
&=& -2\overline{\lambda}
\nonumber \\ 
&& -\dfrac{1}{16\pi^2}\left[ \ln\left( \dfrac{m_{\ell}^2}{\mu^2}\right) 
-\dfrac{M_K^2}{M_K^2-m_{\ell}^2}\,\ln\left( \dfrac{m_{\ell}^2}{M_K^2}\right) \right] \,.
\end{eqnarray} 

\begin{eqnarray}
B(m_{\ell}^2,0,M_K^2)
&=& -2\overline{\lambda}
+\dfrac{1}{16\pi^2}\left[ 1-\ln\left( \dfrac{M_K^2}{\mu^2}\right) \right] 
\nonumber \\ 
&& -\dfrac{1}{16\pi^2}\left( 1-\dfrac{M_K^2}{m_{\ell}^2}\right) \ln\left( 1-\dfrac{m_{\ell}^2}{M_K^2}\right) \,.
\end{eqnarray} 

\begin{eqnarray}
\mathrm{Re}\,B(s_{\pi},M_{\pi}^2,M_{\pi}^2)
&=& \dfrac{1}{M_{\pi}^2}\,A(M_{\pi}^2)
+\dfrac{1}{16\pi^2}\left[ 1-\sigma_{\pi}\ln\left( \dfrac{1+\sigma_{\pi}}{1-\sigma_{\pi}}\right) \right] \,,
\\
\mathrm{Im}\,B(s_{\pi},M_{\pi}^2,M_{\pi}^2)
&=& \dfrac{\sigma_{\pi}}{16\pi}\,,
\end{eqnarray}  
where,
\begin{equation}
\sigma_{\pi}\,=\,\sqrt{1-\dfrac{4M_{\pi}^2}{s_{\pi}}}\,.
\end{equation} 

\begin{eqnarray}
&& B(s_{\pi},M_K^2,M_K^2)\,=\,\dfrac{1}{M_K^2}\,A(M_K^2)+\dfrac{1}{16\pi^2}
\nonumber \\ 
&& -\dfrac{1}{8\pi^2}\left( \dfrac{4M_K^2}{s_{\pi}}-1\right) ^{1/2}\mathrm{arctan}\left( \dfrac{4M_K^2}{s_{\pi}}-1\right) ^{-1/2}\,. 
\end{eqnarray} 

For the following integral, we shall distinguish between two cases:
\begin{itemize}
\item The lepton is an electron.
\begin{eqnarray}
&& B(s_{\pi},M_{\eta}^2,M_{\pi}^2)\,=\,\dfrac{1}{2M_{\eta}^2}\,A(M_{\eta}^2)
+\dfrac{1}{2M_{\pi}^2}\,A(M_{\pi}^2)
\nonumber \\ 
&& +\dfrac{1}{16\pi^2}\left[ 1
-\dfrac{1}{2s_{\pi}}\,(M_{\eta}^2-M_{\pi}^2)\ln\left( \dfrac{M_{\eta}^2}{M_{\pi}^2}\right) \right] 
\nonumber \\ 
&& +\mathrm{If}\left( 4M_{\pi}^2<s_{\pi}<(M_{\eta}-M_{\pi})^2\right) \times
\nonumber \\ 
&& \dfrac{1}{16\pi^2s_{\pi}}\,
\sqrt{(M_{\eta}+M_{\pi})^2-s_{\pi}}\sqrt{(M_{\eta}-M_{\pi})^2-s_{\pi}}\times
\nonumber \\ 
&& \ln\dfrac{\sqrt{(M_{\eta}+M_{\pi})^2-s_{\pi}}+\sqrt{(M_{\eta}-M_{\pi})^2-s_{\pi}}}
{\sqrt{(M_{\eta}+M_{\pi})^2-s_{\pi}}-\sqrt{(M_{\eta}-M_{\pi})^2-s_{\pi}}}
\nonumber \\ 
&& -\mathrm{If}\left( (M_{\eta}-M_{\pi})^2<s_{\pi}<(M_K-m_{\mathrm{e}})^2\right) \times
\nonumber \\ 
&& \dfrac{1}{8\pi^2s_{\pi}}\,
\sqrt{(M_{\eta}+M_{\pi})^2-s_{\pi}}\sqrt{s_{\pi}-(M_{\eta}-M_{\pi})^2}\times
\nonumber \\ 
&& \mathrm{arctan}\,\dfrac{\sqrt{s_{\pi}-(M_{\eta}-M_{\pi})^2}}
{\sqrt{(M_{\eta}+M_{\pi})^2-s_{\pi}}}\,,
\end{eqnarray} 
\item The lepton is a muon.
\begin{eqnarray}
&& B(s_{\pi},M_{\eta}^2,M_{\pi}^2)\,=\,\dfrac{1}{2M_{\eta}^2}\,A(M_{\eta}^2)
+\dfrac{1}{2M_{\pi}^2}\,A(M_{\pi}^2)
\nonumber \\ 
&& +\dfrac{1}{16\pi^2}\left[ 1
-\dfrac{1}{2s_{\pi}}\,(M_{\eta}^2-M_{\pi}^2)\ln\left( \dfrac{M_{\eta}^2}{M_{\pi}^2}\right) \right] 
\nonumber \\
&& +\dfrac{1}{16\pi^2s_{\pi}}\,
\sqrt{(M_{\eta}+M_{\pi})^2-s_{\pi}}\sqrt{(M_{\eta}-M_{\pi})^2-s_{\pi}}\times
\nonumber \\ 
&& \ln\dfrac{\sqrt{(M_{\eta}+M_{\pi})^2-s_{\pi}}+\sqrt{(M_{\eta}-M_{\pi})^2-s_{\pi}}}
{\sqrt{(M_{\eta}+M_{\pi})^2-s_{\pi}}-\sqrt{(M_{\eta}-M_{\pi})^2-s_{\pi}}}\,.
\end{eqnarray} 
\end{itemize} 

\begin{eqnarray}
&& \mathrm{Re}\,B(t_{\pi},m_{\ell}^2,M_{\pi}^2)\,=\,\dfrac{1}{2m_{\ell}^2}\,A(m_{\ell}^2)
+\dfrac{1}{2M_{\pi}^2}\,A(M_{\pi}^2)
\nonumber \\ 
&& +\dfrac{1}{16\pi^2}\left[ 1-\dfrac{1}{2t_{\pi}}\,(m_{\ell}^2-M_{\pi}^2)\ln\left( \dfrac{m_{\ell}^2}{M_{\pi}^2}\right) \right] 
\nonumber \\ 
&& -\dfrac{1}{16\pi^2t_{\pi}}\,
\sqrt{t_{\pi}-(m_{\ell}+M_{\pi})^2}\sqrt{t_{\pi}-(m_{\ell}-M_{\pi})^2}\times
\nonumber \\ 
&& \ln\dfrac{\sqrt{t_{\pi}-(m_{\ell}-M_{\pi})^2}+\sqrt{t_{\pi}-(m_{\ell}+M_{\pi})^2}}
{\sqrt{t_{\pi}-(m_{\ell}-M_{\pi})^2}-\sqrt{t_{\pi}-(m_{\ell}+M_{\pi})^2}}\,, 
\\ 
&& \mathrm{Im}\,B(t_{\pi},m_{\ell}^2,M_{\pi}^2)\,=
\nonumber \\ 
&& \dfrac{1}{16\pi t_{\pi}}\,
\sqrt{t_{\pi}-(m_{\ell}+M_{\pi})^2}\sqrt{t_{\pi}-(m_{\ell}-M_{\pi})^2}\,.
\end{eqnarray} 

\begin{eqnarray}
&& B(t_{\pi},M_{\pi}^2,M_K^2)\,=\,\dfrac{1}{2M_{\pi}^2}\,A(M_{\pi}^2)
+\dfrac{1}{2M_K^2}\,A(M_K^2)
\nonumber \\ 
&& +\dfrac{1}{16\pi^2}\left[ 1-\dfrac{1}{2t_{\pi}}\,(M_{\pi}^2-M_K^2)\ln\left( \dfrac{M_{\pi}^2}{M_K^2}\right) \right] 
\nonumber \\ 
&& +\dfrac{1}{16\pi^2t_{\pi}}\,
\sqrt{(M_{\pi}+M_K)^2-t_{\pi}}\sqrt{(M_{\pi}-M_K)^2-t_{\pi}}\times
\nonumber \\ 
&& \ln\dfrac{\sqrt{(M_{\pi}+M_K)^2-t_{\pi}}+\sqrt{(M_{\pi}-M_K)^2-t_{\pi}}}
{\sqrt{(M_{\pi}+M_K)^2-t_{\pi}}-\sqrt{(M_{\pi}-M_K)^2-t_{\pi}}}\,. 
\end{eqnarray} 

\begin{eqnarray}
&& B(t_{\pi},M_{\eta}^2,M_K^2)\,=\,\dfrac{1}{2M_{\eta}^2}\,A(M_{\eta}^2)
+\dfrac{1}{2M_K^2}\,A(M_K^2)
\nonumber \\ 
&& +\dfrac{1}{16\pi^2}\left[ 1
-\dfrac{1}{2t_{\pi}}\,(M_{\eta}^2-M_K^2)\ln\left( \dfrac{M_{\eta}^2}{M_K^2}\right) \right] 
\nonumber \\ 
&& -\dfrac{1}{8\pi^2t_{\pi}}\,
\sqrt{(M_{\eta}+M_K)^2-t_{\pi}}\sqrt{t_{\pi}-(M_{\eta}-M_K)^2}\times
\nonumber \\ 
&& \mathrm{arctan}\,\dfrac{\sqrt{t_{\pi}-(M_{\eta}-M_K)^2}}
{\sqrt{(M_{\eta}+M_K)^2-t_{\pi}}}\,.
\end{eqnarray}

\subsection{$\tau$-integrals}

These integrals appeared while splitting strong and electromagnetic parts in two-point functions. Their definition is given by equation (\ref{eq:tau_definition}). We are interested in the following particular $\tau$-integrals. 

\begin{eqnarray}
\mathrm{Re}\,\tau (s_{\pi},M_{\pi}^2,M_{\pi}^2)
&=& -\dfrac{2}{s_{\pi}\sigma_{\pi}}\,
\ln\left( \dfrac{1+\sigma_{\pi}}{1-\sigma_{\pi}}\right) \,, 
\\ 
\mathrm{Im}\,\tau (s_{\pi},M_{\pi}^2,M_{\pi}^2)
&=& \dfrac{2\pi}{s_{\pi}\sigma_{\pi}}\,.
\end{eqnarray}

\begin{equation}
\tau (s_{\pi},M_K^2,M_K^2)\,=\,\dfrac{4}{s_{\pi}}\left( \dfrac{4M_K^2}{s_{\pi}}-1\right) ^{-1/2}\mathrm{arctan}\left( \dfrac{4M_K^2}{s_{\pi}}-1\right) ^{-1/2}\,. 
\end{equation} 

\begin{eqnarray}
\tau (t_{\pi},M_{\pi}^2,M_K^2)
&=& \dfrac{2}{\sqrt{(M_{\pi}-M_K)^2-t_{\pi}}\sqrt{(M_{\pi}+M_K)^2-t_{\pi}}}\times
\nonumber \\ 
&& \ln\dfrac{\sqrt{(M_{\pi}+M_K)^2-t_{\pi}}+\sqrt{(M_{\pi}-M_K)^2-t_{\pi}}}
{\sqrt{(M_{\pi}+M_K)^2-t_{\pi}}-\sqrt{(M_{\pi}-M_K)^2-t_{\pi}}}\,.
\end{eqnarray} 

\begin{eqnarray}
\tau (t_{\pi},M_{\eta}^2,M_K^2)
&=& \dfrac{4}{\sqrt{t_{\pi}-(M_{\eta}-M_K)^2}\sqrt{(M_{\eta}+M_K)^2-t_{\pi}}}\times
\nonumber \\ 
&& \mathrm{arctan}\,
\dfrac{\sqrt{t_{\pi}-(M_{\eta}-M_K)^2}}{\sqrt{(M_{\eta}+M_K)^2-t_{\pi}}}\,.
\end{eqnarray} 

\subsection{$C$-integrals}

These are scalar three-point functions whose definition and expressions were given in the appendix of reference~\cite{Nehme:2003bz}. In what follows, we sketch some of the particular cases which we need for the numerical evaluation of Isospin breaking corrections. 
 
\begin{eqnarray}
&& C(m_{\ell}^2,0,m_{\ell}^2,0,m_{\ell}^2,M_K^2)\,=\,
\nonumber \\ 
&& \dfrac{1}{16\pi^2}\left[ \dfrac{1}{m_{\ell}^2}\,\ln\left( 1-\dfrac{m_{\ell}^2}{M_K^2}\right) 
+\dfrac{1}{M_K^2-m_{\ell}^2}\,\ln\left( \dfrac{m_{\ell}^2}{M_K^2}\right)\right] \,.
\end{eqnarray} 

\begin{eqnarray}
&& C(t_{\pi},t_{\pi},0,m_{\ell}^2,M_{\pi}^2,M_K^2)\,=\,
\dfrac{1}{32\pi^2t_{\pi}}\,\dfrac{1}{M_K^2-m_{\ell}^2}\times
\nonumber \\ 
&& \bigg\{\,(M_K^2-M_{\pi}^2+t_{\pi})\ln\left( \dfrac{m_{\ell}^2}{M_K^2}\right) 
\nonumber \\ 
&& +x_0\ln\dfrac{M_K^2-M_{\pi}^2+t_{\pi}+x_0}{M_K^2-M_{\pi}^2+t_{\pi}-x_0}
\nonumber \\ 
&& -x_1\ln\dfrac{M_K^2-M_{\pi}^2+t_{\pi}+x_1}{M_K^2-M_{\pi}^2+t_{\pi}-x_1} 
\nonumber \\ 
&& -x_0\ln\dfrac{(x_0+M_K^2-m_{\ell}^2)^2-\lambda (t_{\pi},m_{\ell}^2,M_{\pi}^2)}
{(x_0-M_K^2+m_{\ell}^2)^2-\lambda (t_{\pi},m_{\ell}^2,M_{\pi}^2)}
\nonumber \\ 
&& +x_1\ln\dfrac{(x_1+M_K^2-m_{\ell}^2)^2-\lambda (t_{\pi},m_{\ell}^2,M_{\pi}^2)}
{(x_1-M_K^2+m_{\ell}^2)^2-\lambda (t_{\pi},m_{\ell}^2,M_{\pi}^2)} 
\nonumber \\ 
&& -(M_K^2-m_{\ell}^2)\ln\dfrac{(x_0+M_K^2-m_{\ell}^2)^2-\lambda (t_{\pi},m_{\ell}^2,M_{\pi}^2)}
{(x_1+M_K^2-m_{\ell}^2)^2-\lambda (t_{\pi},m_{\ell}^2,M_{\pi}^2)} 
\nonumber \\ 
&& -(M_K^2-m_{\ell}^2)\ln\dfrac{(x_0-M_K^2+m_{\ell}^2)^2-\lambda (t_{\pi},m_{\ell}^2,M_{\pi}^2)}
{(x_1-M_K^2+m_{\ell}^2)^2-\lambda (t_{\pi},m_{\ell}^2,M_{\pi}^2)} 
\nonumber \\ 
&& -\lambda^{1/2}(t_{\pi},m_{\ell}^2,M_{\pi}^2)
\ln\dfrac{M_K^2-m_{\ell}^2+x_0+\lambda^{1/2}(t_{\pi},m_{\ell}^2,M_{\pi}^2)}
{M_K^2-m_{\ell}^2+x_0-\lambda^{1/2}(t_{\pi},m_{\ell}^2,M_{\pi}^2)} 
\nonumber \\ 
&& -\lambda^{1/2}(t_{\pi},m_{\ell}^2,M_{\pi}^2)
\ln\dfrac{M_K^2-m_{\ell}^2-x_0+\lambda^{1/2}(t_{\pi},m_{\ell}^2,M_{\pi}^2)}
{M_K^2-m_{\ell}^2-x_0-\lambda^{1/2}(t_{\pi},m_{\ell}^2,M_{\pi}^2)} 
\nonumber \\ 
&& +\lambda^{1/2}(t_{\pi},m_{\ell}^2,M_{\pi}^2)
\ln\dfrac{M_K^2-m_{\ell}^2-x_1+\lambda^{1/2}(t_{\pi},m_{\ell}^2,M_{\pi}^2)}
{M_K^2-m_{\ell}^2-x_1-\lambda^{1/2}(t_{\pi},m_{\ell}^2,M_{\pi}^2)} 
\nonumber \\ 
&& +\lambda^{1/2}(t_{\pi},m_{\ell}^2,M_{\pi}^2)
\ln\dfrac{M_K^2-m_{\ell}^2+x_1+\lambda^{1/2}(t_{\pi},m_{\ell}^2,M_{\pi}^2)}
{M_K^2-m_{\ell}^2+x_1-\lambda^{1/2}(t_{\pi},m_{\ell}^2,M_{\pi}^2)} \,\bigg\}\,, 
\end{eqnarray} 
where, 
\begin{equation}
\lambda^{1/2}(t_{\pi},m_{\ell}^2,M_{\pi}^2)\,=\,
\sqrt{t_{\pi}-(m_{\ell}-M_{\pi})^2}\sqrt{t_{\pi}-(m_{\ell}+M_{\pi})^2}\,,
\end{equation} 
and,
\begin{eqnarray}
x_0
&=& \sqrt{\lambda (t_{\pi},M_{\pi}^2,M_K^2)+4t_{\pi}(M_K^2-m_{\ell}^2)}\,,
\\ 
x_1
&=& \lambda^{1/2}(t_{\pi},M_{\pi}^2,M_K^2)\,=\,
\sqrt{(M_{\pi}-M_K)^2-t_{\pi}}\sqrt{(M_{\pi}+M_K)^2-t_{\pi}}\,.
\end{eqnarray}

\begin{eqnarray}
&& C(M_{\pi}^2,t_{\pi},m_{\ell}^2,0,M_{\pi}^2,M_K^2)\,=\,
\dfrac{1}{16\pi^2}\,\dfrac{1}{m_{\ell}M_{\pi}}\,
\dfrac{\sigma_{\ell \pi}}{1-\sigma_{\ell \pi}^2}\times
\nonumber \\ 
&& \bigg\{\,\ln\left( -\sigma_{\ell \pi}\right) \left[ \ln\left( \dfrac{m_{\ell}M_K}{M_K^2-m_{\ell}^2}\right) +\ln\left( \dfrac{M_{\pi}M_K}{M_K^2-m_{\ell}^2}\right) \right] 
\nonumber \\ 
&& -\dfrac{\pi^2}{6}+\dfrac{1}{2}\,\ln^2\left( \dfrac{m_{\ell}}{M_{\pi}}\right) 
-\ln^2\left( \dfrac{m_{\ell}}{M_K}\right) -\dfrac{1}{2}\,\ln^2\left( -\sigma_{\ell \pi}\right) -\ln^2\left( \sigma_{\pi K}\right) 
\nonumber \\ 
&& -\dfrac{1}{2}\,\ln^2\left( 1-\dfrac{m_{\ell}}{M_{\pi}}\,\sigma_{\ell \pi}\right) 
-\dfrac{1}{2}\,\ln^2\left( 1-\dfrac{M_{\pi}}{m_{\ell}}\,\sigma_{\ell \pi}\right) 
\nonumber \\ 
&& +\dfrac{1}{2}\,\ln^2\left( 1-\dfrac{m_{\ell}}{M_K}\,\dfrac{\sigma_{\ell \pi}}{\sigma_{\pi K}}\right)+\dfrac{1}{2}\,\ln^2\left( 1-\dfrac{M_K}{m_{\ell}}\,\dfrac{\sigma_{\ell \pi}}{\sigma_{\pi K}}\right) 
\nonumber \\ 
&& +\dfrac{1}{2}\,\ln^2\left( 1-\dfrac{m_{\ell}}{M_K}\,\sigma_{\ell \pi}\,\sigma_{\pi K}\right)+\dfrac{1}{2}\,\ln^2\left( 1-\dfrac{M_K}{m_{\ell}}\,\sigma_{\ell \pi}\,\sigma_{\pi K}\right) 
\nonumber \\ 
&& -\mathrm{Li}_2\left( \dfrac{m_{\ell}}{m_{\ell}-M_{\pi}\sigma_{\ell \pi}}\right) 
-\mathrm{Li}_2\left( \dfrac{M_{\pi}}{M_{\pi}-m_{\ell}\sigma_{\ell \pi}}\right) 
\nonumber \\ 
&& +\mathrm{Li}_2\left( \dfrac{m_{\ell}}{m_{\ell}-M_K\sigma_{\ell \pi}\,\sigma_{\pi K}}\right) 
+\mathrm{Li}_2\left( \dfrac{M_K}{M_K-m_{\ell}\sigma_{\ell \pi}\,\sigma_{\pi K}}\right) 
\nonumber \\ 
&& +\mathrm{Li}_2\left( \dfrac{m_{\ell}\sigma_{\pi K}}{m_{\ell}\sigma_{\pi K}-M_K\sigma_{\ell \pi}}\right) 
+\mathrm{Li}_2\left( \dfrac{M_K\sigma_{\pi K}}{M_K\sigma_{\pi K}-m_{\ell}\sigma_{\ell \pi}}\right) \,\bigg\}\,,
\end{eqnarray}  
where,
\begin{eqnarray}
\sigma_{\ell \pi}
&=& \dfrac{\sqrt{t_{\pi}-(m_{\ell}+M_{\pi})^2}-\sqrt{t_{\pi}-(m_{\ell}-M_{\pi})^2}}
{\sqrt{t_{\pi}-(m_{\ell}+M_{\pi})^2}+\sqrt{t_{\pi}-(m_{\ell}-M_{\pi})^2}}\,,
\\ 
\sigma_{\pi K}
&=& \dfrac{\sqrt{(M_{\pi}+M_K)^2-t_{\pi}}-\sqrt{(M_{\pi}-M_K)^2-t_{\pi}}}
{\sqrt{(M_{\pi}+M_K)^2-t_{\pi}}+\sqrt{(M_{\pi}-M_K)^2-t_{\pi}}}\,.
\end{eqnarray}

\begin{eqnarray}
&& \mathrm{Re}\,C(m_{\ell}^2,t_{\pi},M_{\pi}^2,m_{\gamma}^2,m_{\ell}^2,M_{\pi}^2)\,=\,
\dfrac{1}{16\pi^2}\,
\dfrac{1}{m_{\ell}M_{\pi}}\,\dfrac{\sigma_{\ell \pi}}{1-\sigma_{\ell \pi}^2}\times
\nonumber \\ 
&& \bigg\{\,\ln\left( -\sigma_{\ell \pi}\right) \left[ 2\ln\left( 1-\sigma_{\ell \pi}^2\right) -\ln\left( \dfrac{m_{\gamma}^2}{m_{\ell}M_{\pi}}\right) \right] 
\nonumber \\ 
&& +\pi^2+\dfrac{1}{2}\,\ln^2\left( \dfrac{m_{\ell}}{M_{\pi}}\right) 
-\dfrac{1}{2}\,\ln^2\left( -\sigma_{\ell \pi}\right) 
\nonumber \\ 
&& -\dfrac{1}{2}\,\ln^2\left( 1-\dfrac{m_{\ell}}{M_{\pi}}\,\sigma_{\ell \pi}\right) 
-\dfrac{1}{2}\,\ln^2\left( 1-\dfrac{M_{\pi}}{m_{\ell}}\,\sigma_{\ell \pi}\right) 
\nonumber \\ 
&& +\mathrm{Li}_2\left( \sigma_{\ell \pi}^2\right) 
-\mathrm{Li}_2\left( \dfrac{M_{\pi}}{M_{\pi}-m_{\ell}\sigma_{\ell \pi}}\right) 
-\mathrm{Li}_2\left( \dfrac{m_{\ell}}{m_{\ell}-M_{\pi}\sigma_{\ell \pi}}\right) \,\bigg\}\,,
\\ 
&& \mathrm{Im}\,C(m_{\ell}^2,t_{\pi},M_{\pi}^2,m_{\gamma}^2,m_{\ell}^2,M_{\pi}^2)\,=
\nonumber \\ 
&& \dfrac{1}{16\pi}\,
\dfrac{1}{\sqrt{t_{\pi}-(m_{\ell}-M_{\pi})^2}\sqrt{t_{\pi}-(m_{\ell}+M_{\pi})^2}}\bigg [\,\ln (m_{\gamma}^2)
\nonumber \\ 
&& +\ln (t_{\pi})
-2\ln\sqrt{t_{\pi}-(m_{\ell}-M_{\pi})^2}-2\ln\sqrt{t_{\pi}-(m_{\ell}+M_{\pi})^2}\,\bigg ]\,.
\end{eqnarray}

%% file: paper.bbl
\begin{thebibliography}{10}

\bibitem{Nehme:2003bz}
A.~Nehme.
\newblock Isospin breaking in k(l4) decays of the neutral kaon, hep-ph/0311113.
\newblock 2003.

\bibitem{Cabibbo:1965}
N.~Cabibbo and A.~Maksymowicz.
\newblock Angular correlations in $k_{\mathrm{e}4}$ decays and determination of
  low-energy $\pi$-$\pi$ phase shifts.
\newblock {\em Phys. Rev.}, 137:B438--B443, 1965.

\bibitem{Cabibbo:1965E}
N.~Cabibbo and A.~Maksymowicz.
\newblock Angular correlations in $k_{\mathrm{e}4}$ decays and determination of
  low-energy $\pi$-$\pi$ phase shifts.
\newblock {\em Phys. Rev.}, 168:1926, 1968.

\bibitem{Amoros:1999mg}
Gabriel Amoros and Johan Bijnens.
\newblock A parametrization for k+ --> pi+ pi- e+ nu.
\newblock {\em J. Phys.}, G25:1607--1622, 1999.

\bibitem{Santos:2003}
E.~Santos. KTeV~Collaboration.
\newblock An analysis of
  $k_l\longrightarrow\pi^0\pi^{\mp}\mathrm{e}^{\pm}\nu_{\mathrm{e}}(\overline{%
\nu}_{\mathrm{e}})$.
\newblock Talk given at the apr03 meeting of the american physical society,
  Philadelphia, April 5-8, 2003.

\bibitem{Makoff:1993xb}
G.~Makoff et~al.
\newblock Study of the decay k(l) $\to$ pi+- pi0 e-+ anti-neutrino (neutrino).
\newblock {\em Phys. Rev. Lett.}, 70:1591--1594, 1993.

\bibitem{Makoff:1993xbE}
G.~Makoff et~al.
\newblock Study of the decay k(l) $\to$ pi+- pi0 e-+ anti-neutrino (neutrino).
\newblock {\em Phys. Rev. Lett.}, 75:2069, 1995.

\bibitem{Stern:private}
J.~Stern.
\newblock private communication.

\bibitem{Melrose:1965kb}
D.~B. Melrose.
\newblock Reduction of feynman diagrams.
\newblock {\em Nuovo Cim.}, 40:181--213, 1965.

\bibitem{Hagiwara:2002fs}
K.~Hagiwara et~al.
\newblock Review of particle physics.
\newblock {\em Phys. Rev.}, D66:010001, 2002.

\bibitem{Leutwyler:1996qg}
H.~Leutwyler.
\newblock The ratios of the light quark masses.
\newblock {\em Phys. Lett.}, B378:313--318, 1996.

\bibitem{Gasser:1985gg}
J.~Gasser and H.~Leutwyler.
\newblock Chiral perturbation theory: Expansions in the mass of the strange
  quark.
\newblock {\em Nucl. Phys.}, B250:465, 1985.

\bibitem{Rosselet:1977pu}
L.~Rosselet et~al.
\newblock Experimental study of 30,000 k(e4) decays.
\newblock {\em Phys. Rev.}, D15:574, 1977.

\bibitem{Amoros:1999qq}
G.~Amoros, J.~Bijnens, and P.~Talavera.
\newblock Low energy constants from k(l4) form-factors.
\newblock {\em Phys. Lett.}, B480:71--76, 2000.

\bibitem{Bijnens:2002hp}
Johan Bijnens and Pere Talavera.
\newblock Pion and kaon electromagnetic form factors.
\newblock {\em JHEP}, 03:046, 2002.

\bibitem{Buettiker:2003pp}
P.~Buettiker, S.~Descotes-Genon, and B.~Moussallam.
\newblock A re-analysis of pik scattering a la roy and steiner, hep-ph/0310283.
\newblock 2003.

\bibitem{Moussallam:1999aq}
Bachir Moussallam.
\newblock N(f) dependence of the quark condensate from a chiral sum rule.
\newblock {\em Eur. Phys. J.}, C14:111--122, 2000.

\bibitem{Baur:1997ya}
Robert Baur and Res Urech.
\newblock Resonance contributions to the electromagnetic low energy constants
  of chiral perturbation theory.
\newblock {\em Nucl. Phys.}, B499:319--348, 1997.

\end{thebibliography}
